\pdfoutput=1
\documentclass[prd,aps,nofootinbib,twocolumn,superscriptaddress,preprintnumbers,balancelastpage,longbibliography]{revtex4-1}
\usepackage{graphicx}	
\usepackage{amsmath}	
\usepackage{dcolumn}
\usepackage{bm}
\usepackage{graphics}
\usepackage{afterpage}
\usepackage{float}
\usepackage{subfigure}
\usepackage{rotating}
\usepackage{multirow}
\usepackage{tabularx}
\usepackage{booktabs}
\usepackage{multirow}
\usepackage{fancyhdr}
\usepackage{hyperref}
\usepackage[utf8]{inputenc}
\usepackage{theorem}
\usepackage{moreverb}
\usepackage{euscript}
\usepackage{psfrag}
\usepackage{slashed}
\usepackage{mathtools}
\usepackage{makecell}
\usepackage[flushleft]{threeparttable}
\usepackage{adjustbox}
\let\vec\mathbf
\usepackage{dcolumn}
\usepackage{bm}
\usepackage[mathlines]{lineno}
\usepackage{lettrine}
\usepackage[dvipsnames]{xcolor}

\DeclareUnicodeCharacter{2212}{-}

\input Zallman.fd

\LettrineTextFont{\itshape}
\hypersetup{
     colorlinks   = true,
     citecolor    = orange,
     urlcolor     = orange,
     linkcolor    = orange
}

\usepackage{listings}
\usepackage{color,xcolor}

\newcolumntype{C}[1]{>{\centering\let\newline\\\arraybackslash\hspace{0pt}}m{#1}}

\begin{document}

\preprint{SLAC-PUB-17722}
\preprint{LTH-1328}

\title{Evaporation Barrier for Dark Matter in Celestial Bodies}

\author{Javier F. Acevedo}
\thanks{ \href{mailto:jfacev@slac.stanford.edu}{jfacev@slac.stanford.edu}; \href{https://orcid.org/0000-0003-3666-0951}{0000-0003-3666-0951}}
\affiliation{Particle Theory Group, SLAC National Accelerator Laboratory, Stanford, CA 94035, USA}

\author{Rebecca K. Leane}
\thanks{\href{mailto:rleane@slac.stanford.edu}{rleane@slac.stanford.edu}; \href{http://orcid.org/0000-0002-1287-8780}{0000-0002-1287-8780}}
\affiliation{Particle Theory Group, SLAC National Accelerator Laboratory, Stanford, CA 94035, USA}
\affiliation{Kavli Institute for Particle Astrophysics and Cosmology, Stanford University, Stanford, CA 94035, USA}

\author{Juri Smirnov}
\thanks{ \href{mailto:juri.smirnov@liverpool.ac.uk}{juri.smirnov@liverpool.ac.uk}; \href{http://orcid.org/0000-0002-3082-0929}{0000-0002-3082-0929}}
\affiliation{Department of Mathematical Sciences, University of Liverpool,
Liverpool, L69 7ZL, United Kingdom}

\date{\today}

\newcommand{\mk}[1]{{\bf #1}}
\newcommand{\om}[1]{\textcolor{red}{#1}}
\newcommand{\sh}[1]{\textcolor{blue}{#1}}
\newcommand{\aap}{Astronomy and Astrophysics}
\newcommand{\mnras}{Monthly Notices of the RAS}

\begin{abstract}
The minimum testable dark matter (DM) mass for almost all DM signatures in celestial bodies is determined by the rate at which DM evaporates. DM evaporation has previously been calculated assuming a competition between the gravitational potential of the object, and thermal kicks from the celestial-body matter. We point out a new effect, where mediators with a range larger than the interparticle spacing induce a force proportional to the density gradient of celestial objects, forming an \textit{evaporation barrier} for the DM. This effect can be so significant that evaporation does not occur even for sub-MeV DM, in stark contrast to previous calculations. This opens up a wide range of new light DM searches, many orders of magnitude in DM mass below the sensitivity of direct detection.
\end{abstract}
\maketitle

\lettrine{D}{ark} matter (DM) incident on celestial bodies provides a wealth of opportunities for its discovery. One of our most powerful probes of DM comes from direct detection, which tests the rate at which incoming DM scatters with Standard Model (SM) detectors on the celestial body closest to us -- Earth. Direct detection experiments have increasingly constrained GeV-scale DM~\cite{Aalbers:2022dzr,LZ:2022ufs}. However, as the DM mass becomes lighter, its momentum decreases, decreasing the probability it produces a recoil above detector thresholds, and therefore weakening the sensitivity of direct detection. This has motivated a range of novel techniques and detectors to push to lower thresholds, and test the DM-SM scattering rate on the Earth at sub-GeV masses~\cite{Hochberg:2022apz,Hochberg:2015pha, Hochberg:2015fth, Hochberg:2021pkt, Hochberg:2021ymx,Hochberg:2019cyy,Chiles:2021gxk,Schutz:2016tid, Knapen:2016cue, Caputo:2019cyg,Griffin:2018bjn, Knapen:2017ekk, Hochberg:2017wce, Geilhufe:2018gry, Geilhufe:2019ndy, Coskuner:2019odd,Das:2022srn}.

Another promising avenue to test the DM-SM scattering rate at sub-GeV DM masses is to perform new searches with a range of celestial objects. The general idea is that DM can scatter, lose energy, and become gravitationally captured by celestial bodies, and consequently produce detectable signals. However, for sizable DM signals to appear in the capturing object, the DM must not evaporate, $i.e.$ thermal upscatterings of the DM particles to the escape velocity must be sufficiently infrequent, so that a significant amount of DM can build up over time. The lower mass limit for which this condition is met, $i.e.$ \textit{the evaporation mass}, varies for each object and is effectively a limiting threshold for how low the DM mass can be tested~\cite{Gould:1989hm}. Recently it was pointed out that objects such as exoplanets~\cite{Leane:2020wob}, Jupiter~\cite{Leane:2021tjj}, brown dwarfs~\cite{Leane:2020wob,Leane:2021ihh}, neutron stars, and white dwarfs~\cite{Goldman:1989nd,
Gould:1989gw,
Kouvaris:2007ay,
Bertone:2007ae,
deLavallaz:2010wp,
Kouvaris:2010vv,
McDermott:2011jp,
Kouvaris:2011fi,
Guver:2012ba,
Bramante:2013hn,
Bell:2013xk,
Bramante:2013nma,
Bertoni:2013bsa,
Kouvaris:2010jy,
McCullough:2010ai,
Perez-Garcia:2014dra,
Bramante:2015cua,
Graham:2015apa,
Cermeno:2016olb,
Graham:2018efk,
Acevedo:2019gre,
Janish:2019nkk,
Krall:2017xij,
McKeen:2018xwc,
Baryakhtar:2017dbj,
Raj:2017wrv,
Bell:2018pkk,
Chen:2018ohx,
Garani:2018kkd,
Dasgupta:2019juq,
Hamaguchi:2019oev,
Camargo:2019wou,
Bell:2019pyc,
Acevedo:2019agu,
Joglekar:2019vzy,
Joglekar:2020liw,
Bell:2020jou,
Garani:2020wge,
Leane:2021ihh,Bose:2021yhz,Collier:2022cpr} are all objects which can probe sub-GeV DM, due to their relatively cool cores and high densities. Other objects such as the Sun and Earth allow for a range of powerful new searches~\cite{Batell:2009zp,Pospelov:2007mp,Pospelov:2008jd,Rothstein:2009pm,Chen:2009ab,Schuster:2009au,Schuster:2009fc,Bell_2011,Feng:2015hja,Kouvaris:2010,Feng:2016ijc,Allahverdi:2016fvl,Leane:2017vag,Arina:2017sng,Albert:2018jwh, Albert:2018vcq,Nisa:2019mpb,Niblaeus:2019gjk,Cuoco:2019mlb,Serini:2020yhb,Acevedo:2020gro,Mazziotta:2020foa,Bell:2021pyy,Bose:2021cou}, however they do not have as low evaporation masses in previous calculations~\cite{Gould:1989hm}.

\begin{figure}[t!]
    \centering
     \includegraphics[width=\linewidth]{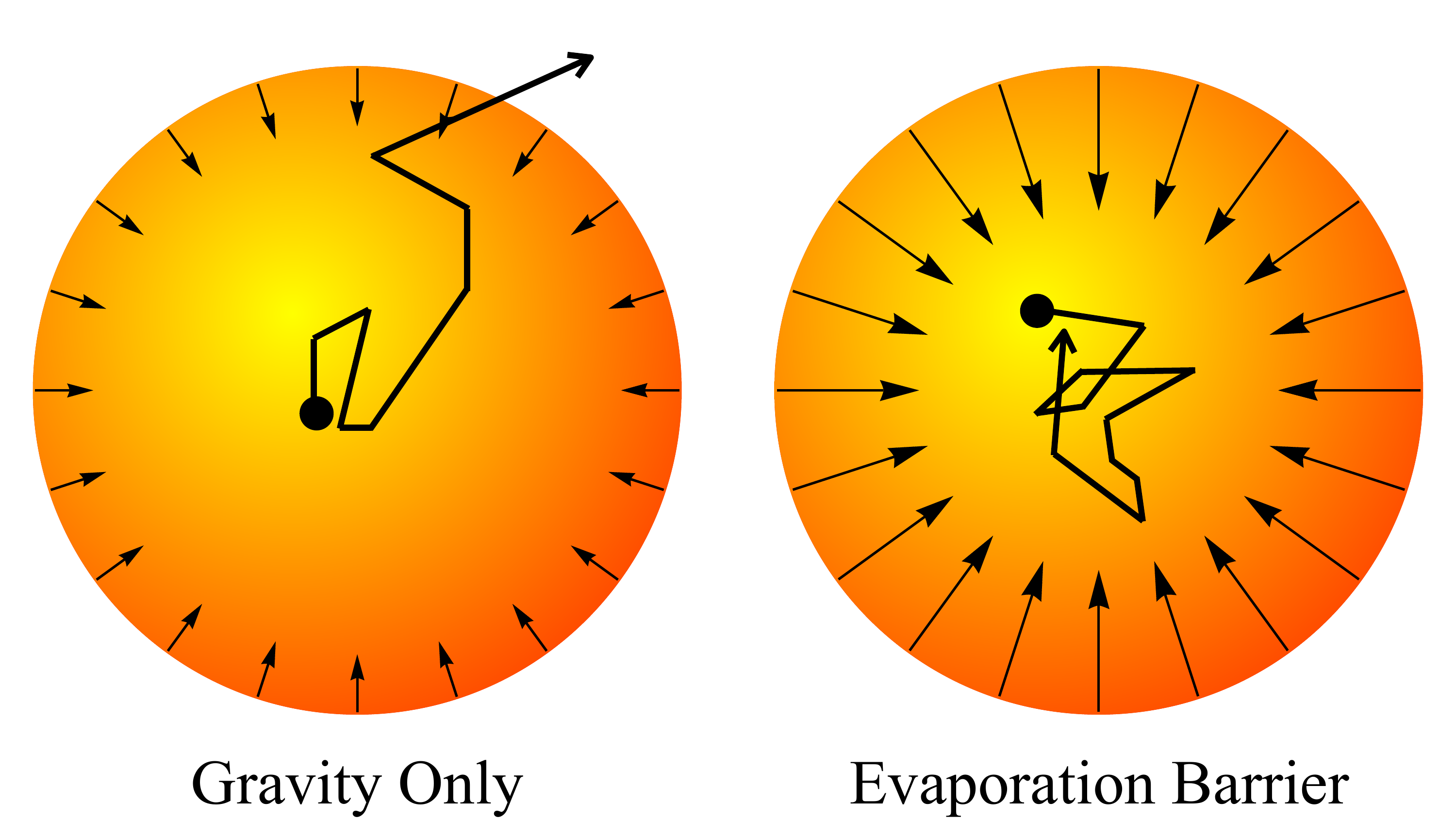}
    \caption{Schematic representation of the evaporation barrier. \textbf{Left:} Gravity only (previous assumption). Evaporation of upscattered light DM particles is not suppressed. \textbf{Right:} The evaporation barrier blocks evaporation (this work).}
    \label{fig:alpha-main}
    \vspace{-3mm}
\end{figure}
In this paper, we show that the evaporation mass for DM in all celestial bodies can be dramatically lower than previously calculated, for a wide range of DM model parameters. This is due to an additional effect that we call the \textit{evaporation barrier}. If the dark sector contains a SM-DM mediator with range larger than the interparticle spacing of the celestial object, the DM will feel an additional force proportional to the celestial-body SM density gradient. This new force can dominate over other effects and allow very light DM particles to be retained in celestial bodies.

The evaporation barrier arises in a wide range of DM parameter space, and is generic in the presence of any attractive long-range force, including models for axions~\cite{Okawa:2021fto}, scalars~\cite{Dimopoulos:1996kp}, and new gauge bosons~\cite{Fayet:1986vz}. We will show that the barrier can be so efficient that even extremely feebly-coupled fields can prevent evaporation. This means that even when a DM-SM contact interaction dominates the scattering processes for capture and evaporation, the mere existence of otherwise decoupled light
fields can be what determines the evaporation rate. As the fundamental nature of DM is completely unknown, considering the effects of a broader range of interaction types better spans the realm of possibilities that can be mapped to new DM searches with celestial bodies.

\section{The Evaporation Barrier}

A fairly generic assumption is that DM can interact with the SM fermions through a dark sector mediator of mass $m_\phi$. Inside a celestial object with SM particle number density $n_{\rm SM}(\vec{r})$, this interaction leads to a potential energy for the DM
\begin{align}
\label{eq:potential}
    \phi_{\rm barrier}(\vec{r}) = - g_{\rm SM} g_\chi \int_V d^3{\vec r'} \, \frac{n_{\rm SM}(\vec{r}') \, e^{- m_\phi | \vec{r}- \vec{r}'| } }{4 \pi |\vec{r} - \vec{r'}|} \, ,
\end{align}
where $V$ is the volume of the object, and $g_{\rm SM}$ and $g_\chi$ are the mediator couplings to SM nucleons and fermionic DM respectively. The effect of this potential becomes relevant once the range of the interaction $r_Y = 1/m_\phi$ becomes larger than the interparticle spacing in the considered object \textit{i.e.} $m_\phi \lesssim \text{keV}$.

The attractive potential has a simple scaling when its range $r_Y$ is such that the SM density varies slowly within the interaction volume. For longer mediator ranges, the field becomes Coulomb-like on the scale of the whole object, and the full solution has to be used (see App.~\ref{app:analyticbarrier}). The simple scaling applies for the objects we consider as long as $m_\phi \gtrsim 10^{-13}$~eV. In this case, the potential can be approximated as $\sim 1/r$ on scales below $r_Y$, and is effectively sourced by the number of particles $N_{\rm eff} \simeq n_{\rm SM}(r) r_Y^3$ within that range,
\begin{align}
\label{eq:approxpotential}
     \phi_{\rm barrier}(r) \simeq - \frac{g_{\rm SM} \, g_\chi N_{\rm eff}  }{r_Y}  \simeq  - \frac{ g_{\rm SM} \, g_\chi n_{\rm SM}(r) }{m_\phi^2}\, ,
\end{align} 
where spherical symmetry is assumed such that $\vec{r}$ is written as $r$. We reproduce this heuristic result in an analytic derivation in App.~\ref{app:analyticbarrier}. For scalar mediators coupled to fermions, or for vector mediators that couple to DM and SM fermions with opposite charge, this potential will generically lead to an attractive force that scales as the density gradient of the celestial body. We call this potential well an evaporation barrier because it serves to pull the DM towards the celestial-body core, preventing DM evaporation.

The evaporation barrier may exist not only due to SM-DM interactions, but also DM self-interactions. Indeed, DM particles accumulated over time can themselves source a field with an analogous scaling relation $\phi(r) \simeq g_\chi^2 n_{\chi}(r)/m_{\phi}^2$. We discuss this possibility, as well as the possible effect of quartic interactions in App.~\ref{app:analyticbarrier}. While the contribution to the evaporation barrier from self-interactions can be comparable to Eq.~\eqref{eq:approxpotential} in some cases, we neglect it in this work and focus only on the DM-SM interaction evaporation barrier contributions. For scalar interactions additional terms will only serve to reinforce the barrier, for vectors the impact will depend on the specific model parameters.

For our purposes of demonstrating the effect of the evaporation barrier, we will assume that contact interactions drive the dynamics other than the evaporation barrier. Such a setup is by no means required; the barrier is a generic feature of attractive long-range forces.\\ 

\section{New Dark Matter Distributions}

To calculate DM evaporation under the presence of the evaporation barrier, we need the DM distribution in celestial objects. We will first address the framework for contact interactions, and discuss alterations due to additional long-range forces shortly. For contact interactions, the distribution is obtained from the Boltzmann Collisional Equation, of which two approximate solutions have been found in the limiting cases of short and long mean free paths.  For a mean free path that is short compared to the scale of temperature variations, local thermal equilibrium (LTE) can be assumed. For the celestial objects considered, this occurs when the cross-section is above $\sim10^{-37}~\rm cm^2 - 10^{-34}~\rm cm^2$. In this case, the radial profile is obtained from~\cite{Gould:1989hm,Leane:2022hkk}
\begin{align}
\label{eq:CEstation}
     \frac{dn_{\chi}(r)}{dr} + n_{\chi}(r)\left[\left(\kappa+1\right) \frac{d\log T(r)}{dr} + \, \frac{f(r)}{T(r)} \right]=0~.
\end{align}
Here $n_{\chi}(r)$ is the DM number density, $\kappa\simeq - 1/[2(1+m_\chi/m_{\rm SM})^{3/2}]$ is a thermal diffusion coefficient with $m_\chi$ the DM mass and $m_{\rm SM}$ the SM target mass~\cite{Leane:2022hkk}, $T(r)$ is the local temperature of the celestial object, and $f(r)$ denotes the total external force, which is where the long-range force enters alongside gravity~\cite{Leane:2022hkk}. Solving this equation for $n_\chi(r)$ provides the DM radial profile in equilibrium, which corresponds to the equilibrium distribution found in Refs.~\cite{Gould:1989hm,Leane:2022hkk}:
\begin{align}
   n_{\chi}^{\rm LTE}(r) = N_0^{\rm LTE} \, \left[ \frac{T(r)}{T(0)}\right]^{3/2-\alpha} \exp{ \left[ - \int_0^r \frac{f(\tilde{r})}{T(\tilde{r})} d\tilde{r} \right] }\, ,
    \label{eq:gould}
\end{align}
where $N_0^{\rm LTE}$ is a constant that normalizes the volume integral of $ n_{\chi}^{\rm LTE}(r)$ to the total number of DM particles in the object $N_\chi$, and where $\alpha \simeq \kappa + 5/2$~\cite{Leane:2022hkk}. We include our new effect as an additional potential energy alongside the gravitational potential energy $\phi_{\rm grav}$,
\begin{equation}
\phi_{\rm tot}(r)=\phi_{\rm grav}(r)+\phi_{\rm barrier}(r),
\end{equation}
such that the net external force due to gravity and the evaporation barrier is
\begin{equation}
f(r)=-\nabla\phi_{\rm tot}(r).
\end{equation}
By contrast, for long mean free paths, an isothermal distribution $n_{\chi}^{\rm iso}(r)$ becomes an increasingly good approximation for the DM number density distribution~\cite{Spergel:1984re}, 
\begin{align}
n_{\chi}^{\rm iso}(r) = N_0^{\rm iso} \, {\exp{\left[-  \frac{\phi_{\rm tot}(r)}{T_{\chi}}\right]}},
\end{align}
where $N_0^{\rm iso}$ is a constant that normalizes the volume integral of $n_{\chi}^{\rm iso}(r)$ to $N_\chi$. In this isothermal approximation, the temperature of the DM particles $T_{\chi}$ is obtained from requiring no net energy flow between the DM and SM baths in a steady state.
In practice, $T_{\chi}$ will be an intermediate value between the central temperature of the celestial body and the temperature at the scale radius $r_\chi$ of the DM distribution. This is approximated by 
\begin{equation}
    r_{\chi} \simeq \left(\frac{3\, T(0)}{2 \pi \,G \, \rho(0)\, m_{\chi}}\right)^{1/2}
\end{equation}
when only the gravitational force is present, where $G$ is the gravitational constant and $\rho(0)$ is the SM density at the core of the object. The correction to this orbital radius due to the additional long-range force does not significantly impact our evaporation rate estimates, as it only shifts the relative weight between the LTE and isothermal distributions within a narrow range of DM-SM cross-sections.

The parameter that differentiates between the isothermal and the LTE regime is the Knudsen number $K = l(0)/r_{\chi}$, which relates the mean free path at the core $l(0)$ to the length-scale of the DM distribution. Here $l(0) = \left(\sum_{i} n_{i}(0) \sigma_{i \chi}\right)^{-1}$ with $i$ running over the constituent species of the celestial-body material, and $\sigma_{i \chi}$ the DM-SM scattering rate. The expression that approximately connects both regimes is~\cite{Bottino:2002pd,Scott:2008ns}
\begin{align}
  n_{\chi}(r) = f(K)\, n_{\chi}^{\rm LTE}(r) + \left[1 - f(K)\right] n_{\chi}^{\rm iso}(r) ,
 \end{align}
where $f(K) = 1 - (1 + (K_0/K)^2)^{-1}$ with $K_0 \approx 0.4$~\cite{Gould:1989hm} for the Sun; we will use the same value as an approximation for all objects. For the full formalism to describe DM distributions for short-range DM in the Sun, Earth, Jupiter, Brown Dwarfs, and Exoplanets, see Ref.~\cite{Leane:2022hkk}.\\

\section{New Dark Matter Evaporation Results}

An analytic formula for the evaporation rate is~\cite{1990ApJ...356..302G},
\begin{align}
  \Gamma_{\rm evap} = 4 \pi \int_0^R \, r^2 \, s(r) \, n_\chi(r) \, \left(\sum_i \xi_i(r) n_i(r) \sigma_{i \chi}\right)\,dr~\,,
  \label{eq:evap}
\end{align}
where in the short mean free path regime, the DM velocity distribution is Maxwellian with $\bar{v}= \sqrt{8 T(r)/(\pi m_\chi)}$, where $\bar{r}$ is the mean DM radius, and $\xi_i(r) = \bar{v}(r) |\hat{\phi}| e^{-|\hat{\phi}|}$, $s(r) = e^{-\tau(r)} \eta(r)$, where the opacity is given by~\cite{1990ApJ...356..302G}
\begin{align}
 \tau(r) = \int_r^R ds \,l(s)^{-1}= \sum_i \int_r^R ds \, n_i(s) \sigma_{i \chi}\, ,
\end{align}
where $l(r)$ is the mean free path of the DM particle, and the trajectory correction is~\cite{1990ApJ...356..302G}
\begin{align}
\eta(r) = \frac{7}{10} \, e^{\frac{3 \tau(r) }{2 \hat{\phi}(r) }} \, \min \left(1,\tau^{-1}(r)\right)    \,.
\end{align}
Here $\hat{\phi}(r)= \phi_{\rm tot}(r)/T(r)$ is the ratio of the escape energy and temperature at radius $r$. The trajectory correction is derived assuming isotropic DM-SM scattering, and using the Chapman-Enskog expansion~\cite{1990ApJ...356..302G}. Eq.~(\ref{eq:evap}) also approximates the evaporation mass value well in the long mean free path regime~\cite{1990ApJ...356..302G}.

\begin{figure}[t]
    \centering
%    \hspace{-5mm}
\includegraphics[width=\linewidth]{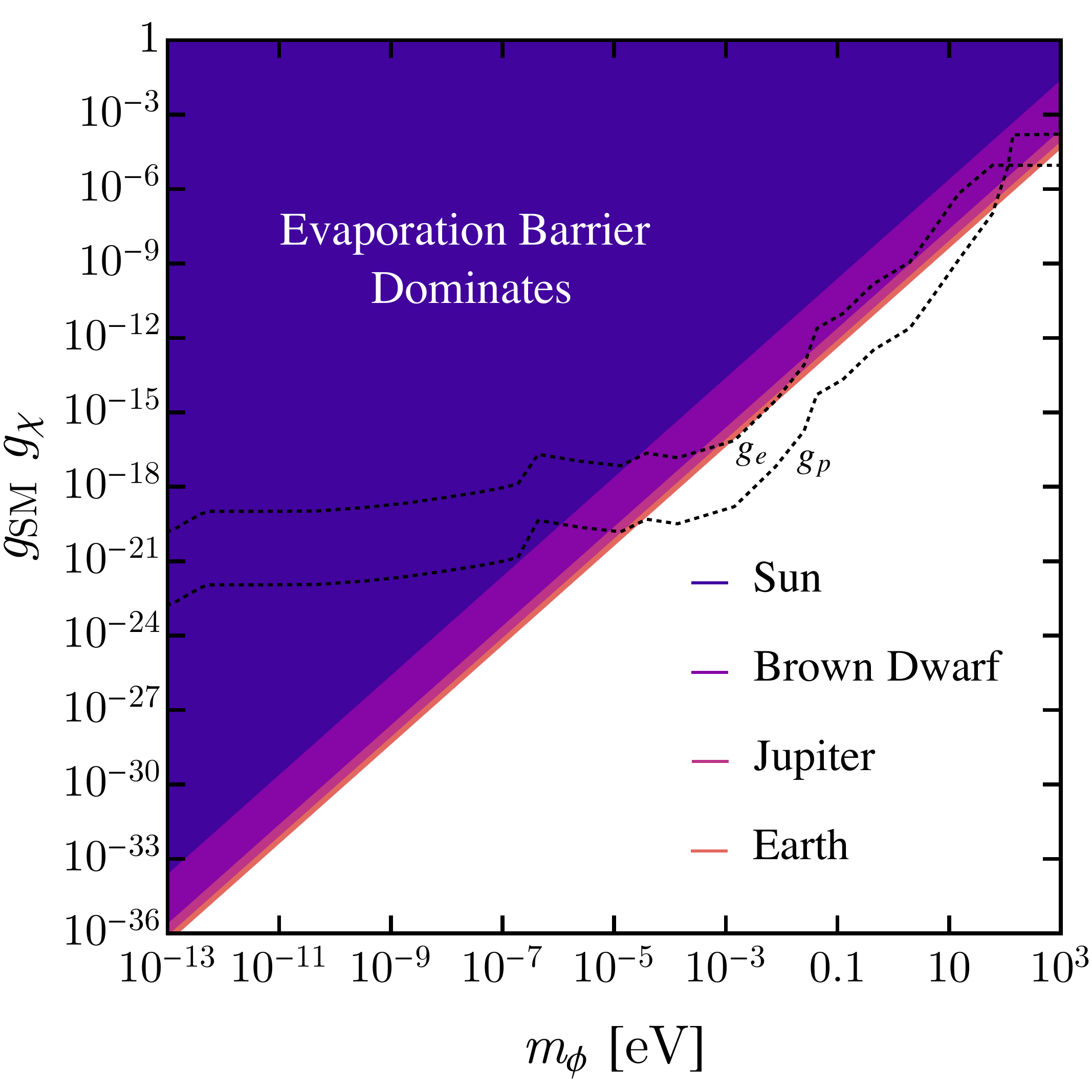}
    \caption{Mediator masses $m_\phi$ and corresponding minimum coupling values $g_{\rm SM} g_{\chi}$ for which the DM evaporation mass is substantially reduced compared to the gravity-only case for each object. The dotted lines correspond to constraints on proton ($g_p$) and electron ($g_e$) couplings from lab experiments.}
    \label{fig:evapthreshold}
\end{figure}

\begin{figure*}[!th]
    \centering
     \includegraphics[width=\columnwidth]{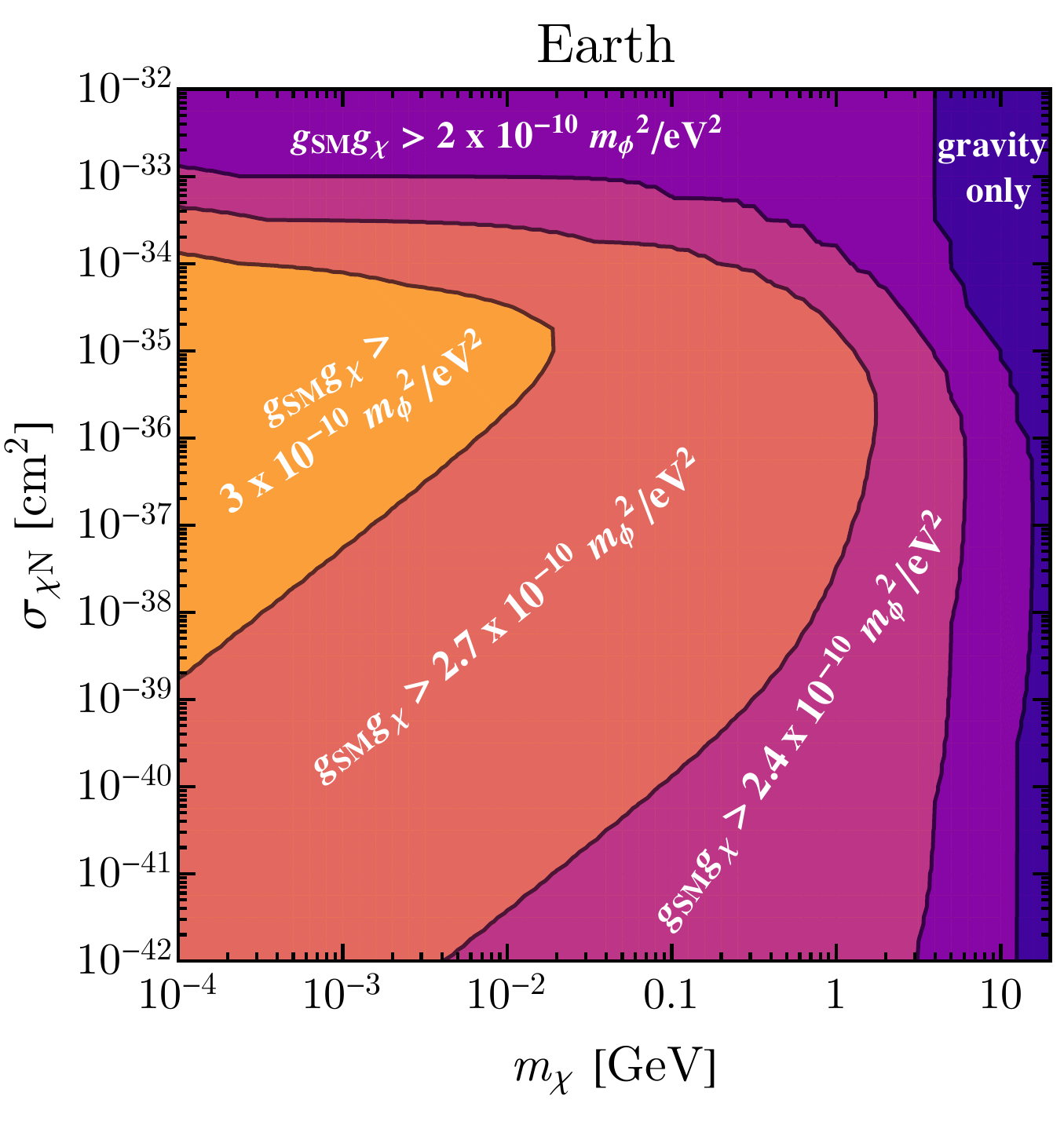}\hspace{5mm}
     \includegraphics[width=\columnwidth]{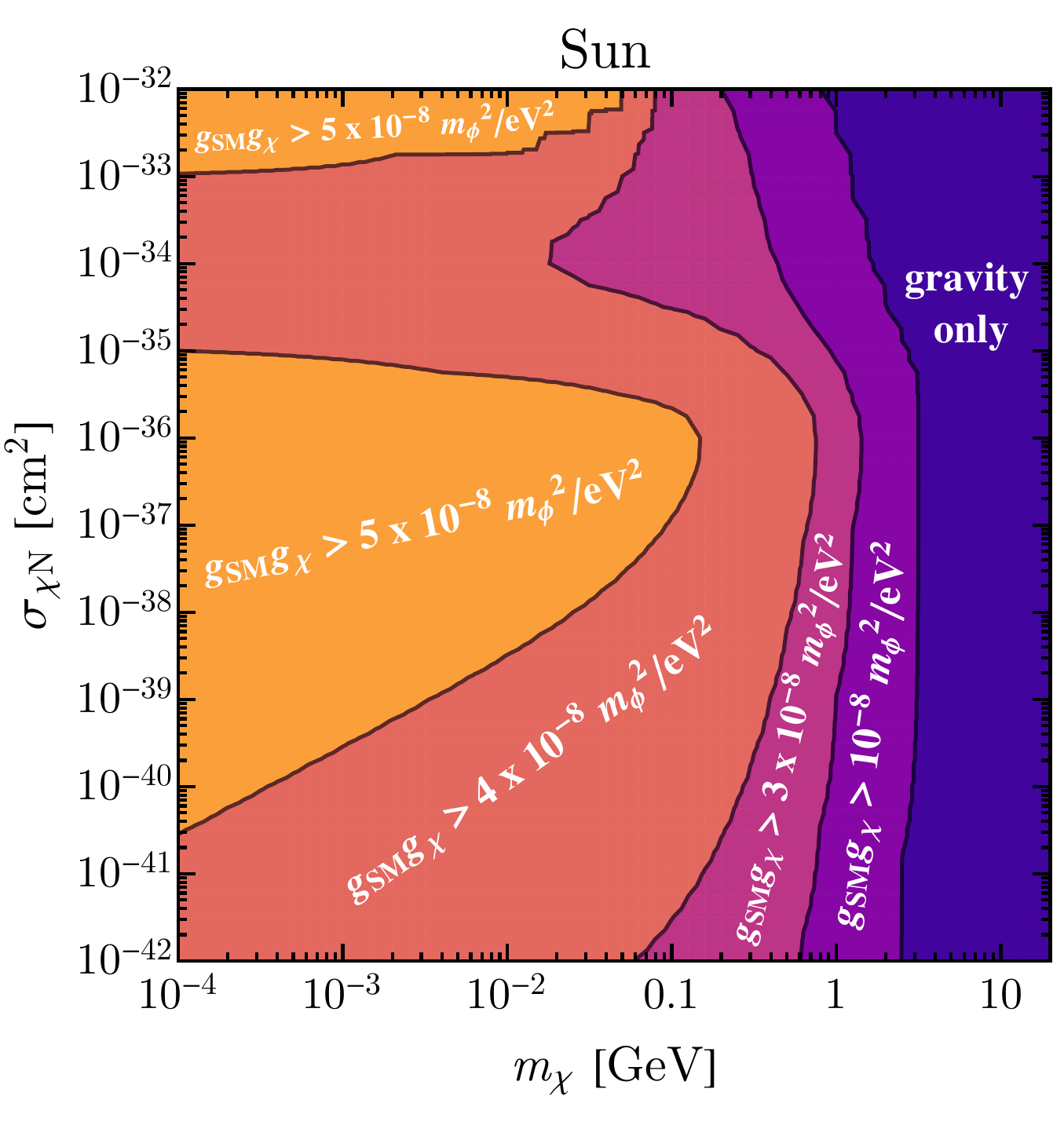}
          \includegraphics[width=\columnwidth]{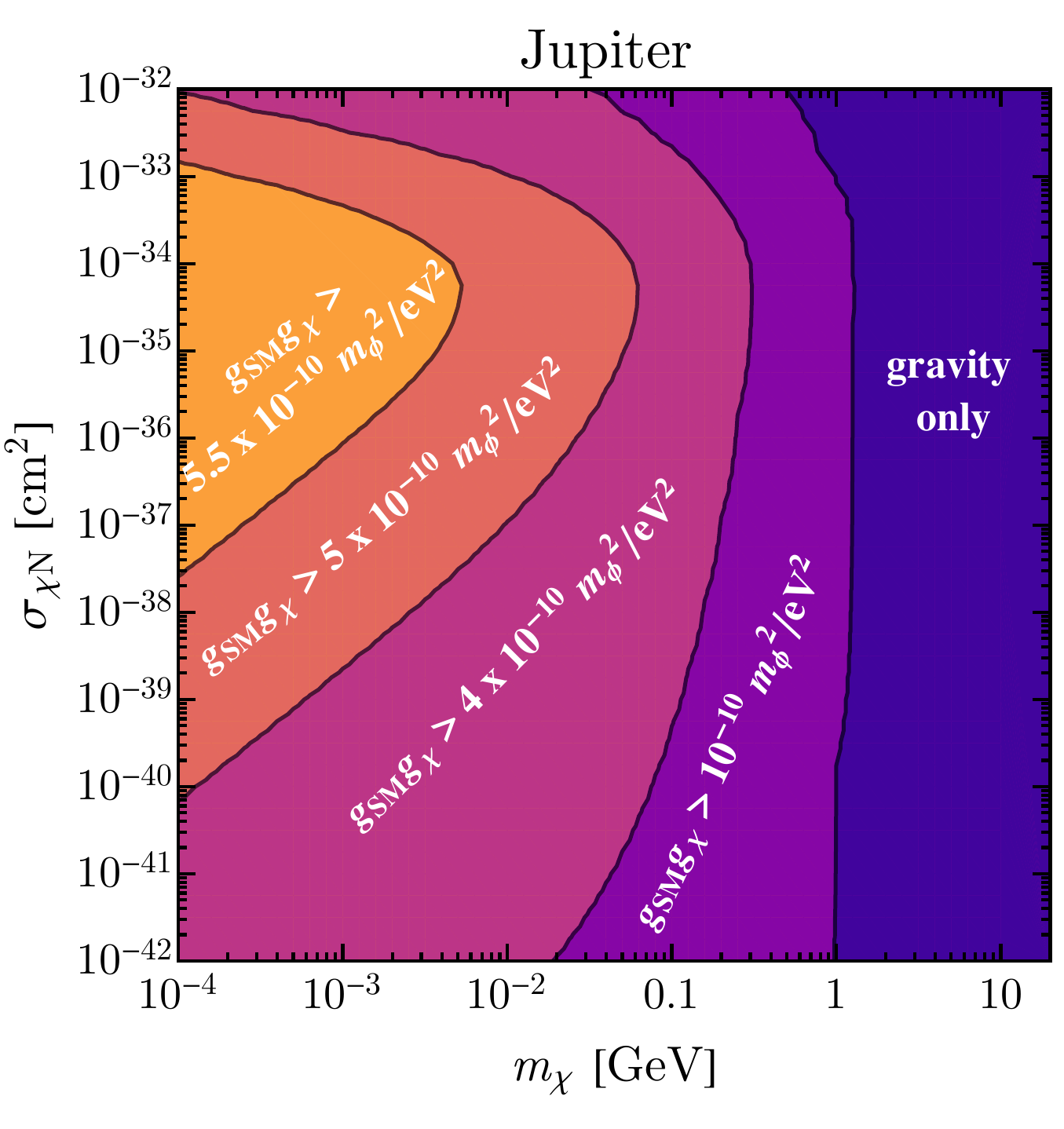}\hspace{5mm}
     \includegraphics[width=\columnwidth]{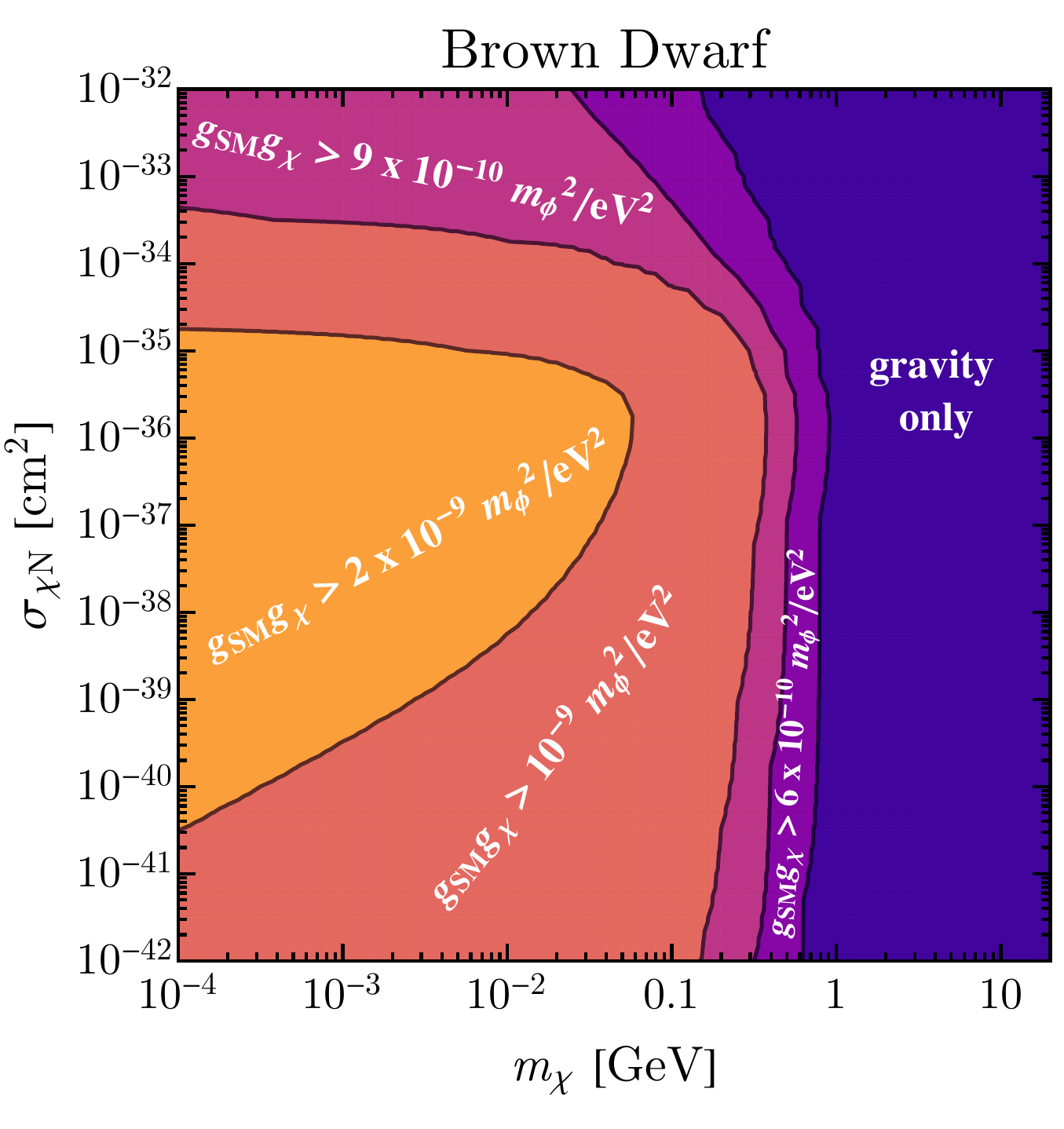}
    \caption{DM evaporation mass contours for example celestial objects assuming an annihilation cross-section $\langle \sigma_{\rm ann} v_{\rm rel} \rangle = 3\times 10^{-26} \, \rm cm^3 / \,s$, and spin-independent DM-SM scattering $\sigma_{\chi N}$. The contour edges are defined by the minimum DM mass $m_\chi$ for which a combination of the mediator coupling to DM $g_\chi$ and SM nucleons $g_{\rm SM}$, and the mediator mass $m_\phi$, lead to DM particles being retained in the celestial object. Right of the ``gravity only" contour corresponds to the only parameter space where DM was retained in previous calculations; that region neglects the evaporation barrier.}
    \label{fig:evapcontours}
\end{figure*}

For a given capture and annihilation cross-section, the evaporation mass is thus approximately defined as the threshold value $m_{\rm evap}$ for which~\cite{1990ApJ...356..302G}
\begin{equation}
    \Gamma_{\rm evap}(m_{\rm evap}) \lesssim t_{\rm eq}^{-1},
\end{equation}
where $t_{\rm eq} \simeq \left(\Gamma_{\rm cap}\Gamma_{\rm ann}\right)^{-1/2}$ is the equilibrium timescale, and $\Gamma_{\rm cap}$ is the capture rate, which we obtain following Ref.~\cite{Leane:2022hkk}. While the capture rate in Eq.~(5) of Ref.~\cite{Leane:2022hkk} only applies in the strong interaction regime, we use the same capture equation even when entering the optically thin regime, and even when the SM target mass is smaller than the DM mass. This choice is made as a simplifying and conservative assumption; in any case, the evaporation rate is only logarithmically sensitive to the capture rate, so the precise value is not very important. As we are assuming the long-range force is decoupled compared to contact interactions, and we furthermore restrict our range to shorter than the celestial-body size, we can neglect long-range capture effects~\cite{Chen:2015uha, Davoudiasl:2017pwe,Davoudiasl:2020ypv,Gaidau:2021vyr,Gresham:2022biw}. The annihilation rate is~\cite{1990ApJ...356..302G}
\begin{align}
    \Gamma_{\rm ann} =  4\pi\,\langle \sigma_{\rm ann} v_{\rm rel} \rangle \, \int_0^R r^2 \,n_{\chi}(r)^2\, dr.
\end{align}
Regardless of the mean free path regime, the presence of long-range forces has important effects to reduce evaporation. First, it increases the required escape energy. Second, it increases the optical depth that any evaporating particle must traverse, since the distribution of DM becomes more concentrated at the center.
 
Figure~\ref{fig:evapthreshold} shows where the evaporation barrier is significant for four objects: the Sun, Earth, Jupiter, and a benchmark 50 Jupiter mass and 10 Gyr Brown Dwarf. In the shaded region, the evaporation barrier dominates, such that the DM evaporation mass will be severely decreased. This region is estimated by requiring $|\phi_{\rm barrier}(r=0)| \gtrsim |\phi_{\rm grav}(r=0)|$, where $\phi_{\rm grav}$ is evaluated at the evaporation mass in the absence of long-range forces. We evaluate this condition at $r = 0$ because $|\phi_{\rm grav}|$ is maximized, so it gives the most conservative estimate on the coupling strength needed for the barrier potential to dominate. Importantly, note the wide range of parameter space for which this is true. Couplings as small as about $10^{-36}$ or larger can have an effect, depending on the mass of the mediator. As the mediator mass increases, its range decreases, which decreases the sum of SM targets available for interactions. Therefore, larger couplings are required for the evaporation barrier to have a significant effect for shorter-range mediators. Depending on the DM model realization, the parameter space is subject to some constraints, we show the electron and proton lab constraints in Fig.~\ref{fig:evapthreshold}. See App.~\ref{app:models} for more details and some applicable model examples. 

The evaporation barrier also exists for mediator masses $m_\phi\lesssim10^{-13}$~eV, but requires use of the full solution, see App.~\ref{app:analyticbarrier}. We model the SM interior of these objects as per Ref.~\cite{Leane:2022hkk}, although we do not include Earth's atmosphere. We have so far not discussed higher order corrections such as cubic and quartic interactions, which can be relevant for our barrier effect, these are discussed in App.~\ref{app:analyticbarrier}. From the point of view of the low energy theory that we are interested in, the self interactions are renormalized quantities, and their value is a free observational parameter. The logarithmic running of the couplings, however, induces contributions to self interactions at larger energy scales. In particular in the case of the scalar potential physical consistency requires that in the UV theory no vacuum instability is triggered. Possible UV embeddings should certainly avoid vacuum decay, but are beyond the scope of our discussion at this point.

Figure~\ref{fig:evapcontours} shows DM evaporation contours with the evaporation barrier included. Regions that are shaded correspond to DM being retained in the celestial body (i.e. there is no significant evaporation), for the condition labeled on the contour. The previous classic DM evaporation framework of Ref.~\cite{Gould:1989hm}, widely used in the literature (see e.g. Ref.~\cite{Garani:2021feo}) is shown as the ``gravity only'' region. The regions below this in DM mass show the impact of the evaporation barrier -- the DM evaporation mass can be extremely sensitive to even tiny couplings of the mediator. We discuss the behavior of these contours further in App.~\ref{app:escapes}. In the case of a vector mediator, neglecting self-interactions, only DM of one charge will be retained, similar to an asymmetric DM scenario. However, this only leads to an $\mathcal{O}(1)$ increase in the evaporation mass. We re-iterate our simplifying assumptions in making these contours: no self-interaction effects, and we assume that the light mediator is de-coupled from the capture, annihilation, and evaporation scattering events, which are instead mediated by contact interactions. These assumptions are by no means required for the barrier to have an effect; they are only simplifying assumptions to explicitly demonstrate the impact of the barrier. We have also assumed only free particle interactions for scattering; for DM masses comparable to or lighter than the edge of our shown DM mass range of 100 keV, collective effects can also be relevant~\cite{DeRocco:2022rze}. We expect however that qualitatively similar results would be obtained in that regime.\\

\section{Implications and Outlook}
The fundamental particle model for DM is unfortunately \textit{completely} unknown. In light of the span of vast possibilities for its nature, the community should perform DM searches in a manner as model-independent as possible. A wide range of celestial bodies have been shown to be excellent targets for DM detection, with their lower DM-mass reaches truncated by the evaporation process. We have shown for the first time that the DM mass at which DM evaporates in celestial objects is, in reality, severely model-dependent. Simple, widespread, and well-motivated extensions of the SM provide forces that can prevent lighter DM from evaporating, allowing sizable DM populations in celestial objects. This leads to sensitivities many orders of magnitude in DM mass below thresholds of other probes of the DM-SM scattering rate, such as direct detection experiments.

We have focused on the effects of the evaporation barrier for some of the most promising targets for DM searches: the Sun, the Earth, Jupiter, and Brown Dwarfs. For each case, we have found dramatic multi-order-of-magnitude reductions are possible for the DM evaporation mass, even for extremely feeble interactions. Our results open up new searches for sub-GeV and ultra-light DM in celestial objects, and significantly expand DM implications for existing searches. The implications include:
\begin{itemize}
    \item Solar neutrino searches from DM annihilation at experiments such as Super-K~\cite{Super-Kamiokande:2004pou,Super-Kamiokande:2015xms}, IceCube~\cite{IceCube:2012ugg,IceCube:2016yoy}, Antares~\cite{ANTARES:2016xuh}, should not cut their search sensitivity at the previously calculated DM evaporation mass of $\sim4$ GeV for the Sun, which we have shown is a highly model-dependent value. Searches for neutrinos from DM annihilation in the Earth at SuperK~\cite{Mijakowski_2020}, Antares~\cite{ANTARES:2016bxz}, IceCube~\cite{IceCube:2016aga,IceCube:2020wxa,IceCube:2021eqb} should not cut their sensitivity at the previously calculated $\sim10$~GeV for the Earth. This opens substantial additional DM parameter space in neutrino searches.
    
    \item Searches for gamma rays from DM annihilation in the Sun~\cite{Meade:2009mu,Batell:2009zp,Schuster:2009au,Leane:2017vag,Albert:2018jwh}, Brown Dwarfs~\cite{Leane:2021ihh,Bhattacharjee:2022lts} and Jupiter~\cite{Leane:2021tjj} can likewise extend to at least sub-MeV DM mass, depending on the DM model. 
    
    \item Searches for electron final states in the Sun~\cite{Schuster:2009fc,Schuster:2009au,FermiLAT:2011ozd,Feng:2016ijc}, or Jupiter~\cite{Li:2022wix} can be extended to at least sub-MeV DM masses.
    
    \item Exoplanet or brown dwarf heating~\cite{Leane:2020wob} can extend to at least sub-MeV DM masses.
    
    \item Solar reflection constraints, as pointed out in Refs.~\cite{Kouvaris:2015nsa,An:2017ojc,Emken:2017hnp,An:2021qdl}, may now disappear for sufficiently coupled light mediators, and need reevaluation. Solar reflection is the idea that DM can scatter with the Sun, gain energy from the solar temperature, evaporate, and travel to Earth with higher energies, making sub-GeV DM sufficiently energetic to overcome direct detection thresholds. However, now that evaporation can be blocked by the barrier, DM may remain in the Sun and not be detectable by this method. This would remove the intrinsic component of Galactic DM expected from the Sun for sub-GeV DM, as usually shown as an irreducible constraint on the direct detection parameter space~\cite{SENSEI:2020dpa,Du:2022dxf,Essig:2022dfa}.
    
     \item The different DM distributions lead to new observables. For example, this may affect the solar abundance problem, which has not yet been solved, and presents a $6\sigma$ discrepancy between theory and experiment~\cite{Asplund_2009, Serenelli_2009, Bergemann_2014,Frandsen:2010yj,Taoso:2010tg,Cumberbatch:2010hh,Vincent:2014jia,Vincent:2015gqa,Vincent:2016dcp,Banks:2021sba}. Further, DM effects on stellar evolution and astroseismology may also be different than previously studied~\cite{PhysRevLett.108.061301, refId0, 10.1093mnrasstab865, 2019asym, Rato:2021tfc}. In addition, this means that surface DM abundances can be different to previously calculated~\cite{Leane:2022hkk}, but low-threshold direct detection of the thermalized DM population~\cite{Das:2022srn} may extend to lower DM masses.
\end{itemize}

We recommend that experimental collaborations, and theorists using data, analyze celestial-body data as low as experimental thresholds allow, rather than as low as one particular DM model's result for the DM evaporation mass. After all, the span of DM possibilities is vast, and in the end, anything could be realized by nature. Wherever DM is hiding, we don't want to miss it.\\

\section*{Acknowledgments} We thank J. Beacom, C. Cappiello, D. Cereskaite, D. Curtin, T. Emken,  R. Garani, T. Linden, and A. Vincent, for helpful discussions and comments. JFA and RKL are supported in part by the U.S. Department of Energy under Contract DE-AC02-76SF00515. 

%\newpage
%\maketitle
\onecolumngrid
%\begin{center}
%\textbf{\large Evaporation Barrier for Dark Matter in Celestial Bodies}

%\vspace{0.05in}
%{ \it \large Supplemental Material}\\ 
%\vspace{0.05in}
%{Javier F. Acevedo, Rebecca K. Leane, and Juri Smirnov}
%\end{center}
%\onecolumngrid
%\setcounter{equation}{0}
%\setcounter{figure}{0}
%\setcounter{section}{0}
%\setcounter{table}{0}
%\setcounter{page}{1}
%\makeatletter
%\renewcommand{\theequation}{S\arabic{equation}}
%\renewcommand{\thefigure}{S\arabic{figure}}
%\renewcommand{\thetable}{S\arabic{table}}

%\tableofcontents
\appendix

\section{Analytic Solution for the Evaporation Barrier}
\label{app:analyticbarrier}
The evaporation barrier is the potential energy a DM particle has due to the SM density $n_{\rm SM}$ of the celestial object. We express this as $\phi_{\rm barrier}(r) = g_{\chi} \phi(r)$, where the field $\phi(r)$ follows from the inhomogeneous time-independent Klein-Gordon equation
\begin{equation}
    (\mathbf{\nabla}^2 - m_{\phi}^2)\phi(\mathbf{r}) = g_{\rm SM} \, n_{\rm SM}(\mathbf{r})~,
    \label{eq:app-a-main}
\end{equation}
and $g_{\rm SM}$ and $g_\chi$ is the SM and DM coupling to new light sectors respectively. This coupling is in principle renormalized so that it accounts for potential loop-order corrections. Depending on the specific model realization, the field $\phi$ could be a scalar or the time-like component of a time-independent vector field. Since the evolution of the celestial bodies takes place on much slower timescales than the DM dynamics, this time-independence is justified. The set-up is analogous to the electrostatic solutions in classical electrodynamics, and we will refer to this regime as \textit{static}. We proceed to derive an analytic solution for this equation given an arbitrary spherically-symmetric density profile. In particular, we will show below that, in the limit that the mediator mass $m_\phi$ is heavy compared to the inverse of the size $R$ of the celestial object, this external potential becomes proportional to the local density of SM particles. 
 
 In spherical coordinates, we impose as boundary conditions $\frac{d\phi}{dr}|_{r=0}=0$ and $\phi(r>R) = \phi(R) (R/r)\exp(-m_{\phi}(r-R))$. The central field value $\phi(0)$, as well as the value at the boundary $\phi(R)$, are determined below from requiring continuity and differentiability. Writing Eq.~\eqref{eq:app-a-main} explicitly in spherical coordinates, and introducing the change of variables $\psi(r) = r \, \phi(r)$, we get
\begin{equation}
    \frac{d^2\psi}{dr^2}-m_{\phi}^2 \psi(r) = g_{\rm SM} \, r \, n_{\rm SM}(r)~.
    \label{eq:app-a-aux}
\end{equation}
In terms of the auxiliary field $\psi$, the boundary conditions read $\psi(0)=0$ and $\psi'(0)=\phi(0)$. The solution to the above equation, for the boundary conditions we have imposed here, is
\begin{equation}
    \psi_{\rm int}(r)=\psi_1(r)+g_{\rm SM} \int_0^r \psi_2(r-s)\ n_{\rm SM}(s)\, s \, ds\,,
    \label{eq:app-a-aux2}
\end{equation}
where $\psi_1$ is the solution to the homogeneous problem ($i.e.$ Eq.~\eqref{eq:app-a-aux} without the source term) with boundary conditions $\psi_1(0)=0$ and $\psi_1'(0)=\phi(0)$, while similarly $\psi_2(r)$ is the solution to the homogeneous problem but with boundary conditions $\psi_2(0)=0$ and $\psi_2'(0)=1$. These are
\begin{equation}
    \psi_1(r)=\phi(0)\sinh(m_{\phi}r)~,
\end{equation}
and
\begin{equation}
    \psi_2(r)=\sinh(m_{\phi}r)~.
\end{equation}
The solution for the interior of the celestial object must be smoothly connected to the solution outside, which will simply scale as $\phi_{\rm ext}(r) = \phi(R) (R/r) \exp\left(m_{\phi}(R-r)\right)$. Both $\phi(0)$ and $\phi(R)$ are then determined from this boundary condition. Switching back to $\phi(r)=\psi(r)/r$, this leads to a system of linear equations for the pair $\left(\phi(0),\phi(R)\right)$ of the form
\[
\begin{bmatrix}
\sinh(m_{\phi}R) & -m_{\phi}R \\
\cosh(m_{\phi}R) & m_{\phi}R 
\end{bmatrix}
\begin{bmatrix}
\phi(0) \\ \phi(R) 
\end{bmatrix}
=
\begin{bmatrix}
a \\ b
\end{bmatrix}
\]
where $a$ and $b$ are given by
\begin{equation}
    a = - g_{\rm SM} \int_0^R \sinh\left(m_{\phi}(R-s)\right) \, s \, n_{\rm SM}(s) \, ds\,,
\end{equation}
\begin{align}
    b & = - g_{\rm SM} \frac{d}{dr} \left[\int_0^r \sinh\left(m_{\phi}(r-s)\right) \, s \,  n_{\rm SM}(s) \, ds\right]_{r=R} \\ &  = - g_{\rm SM} \int_0^R \cosh\left(m_{\phi}(R-s)\right) \, s \,  n_{\rm SM}(s) \, ds\,. \nonumber
\end{align}

The system above is non-singular and can be easily solved by inverting the coefficient matrix, which yields a solution for $\phi(0)$ of the form
\begin{equation}
    \phi(0)= - g_{\rm SM} \int_0^R \exp(-m_{\phi}s) \ s \ n_{\rm SM}(s) \ ds\,.
\end{equation}
The formal solution to Eq.~\eqref{eq:app-a-main} with the boundary conditions we imposed thus reads
\begin{equation}
    \phi_{\rm int}(r)=\phi(0) \left(\frac{\sinh(m_{\phi}r)}{m_{\phi}r}\right) + g_{\rm SM} \int_0^r \frac{\sinh\left(m_{\phi}(r-s)\right)}{m_{\phi}r} s  \ n_{\rm SM}(s) \ ds\,,
    \label{eq:app-a-mainsol-int}
\end{equation}
and
\begin{equation}
    \phi_{\rm ext}(r)=\phi(R) \left(\frac{R}{r}\right) \exp\left(-m_{\phi}(r-R)\right)\,.
    \label{eq:app-a-mainsol-ext}
\end{equation}
In the limit that the range of the interaction is short compared to the length scale of density variation, \textit{i.e.} $m_{\phi}^{-1} \ll n_{\rm SM}(r)/n_{\rm SM}'(r)$ across the celestial object, a simple scaling relation for the field can be obtained. This can be seen upon expanding the hyperbolic functions in Eq.~\eqref{eq:app-a-mainsol-int}:
\begin{align}
    \phi_{\rm int}(r) & = - g_{\rm SM} \int_r^R \frac{\exp(m_{\phi}(r-s))}{2m_{\phi}r} \, s \, n_{\rm SM}(s) \ ds \nonumber \\ & - g_{\rm SM} \int_0^r \frac{\exp(-m_{\phi}(r-s))}{2m_{\phi}r} \, s \, n_{\rm SM}(s) \, ds \\ & + g_{\rm SM} \int_0^R \frac{\exp(-m_{\phi}(r+s))}{2m_{\phi}r} \, s \, n_{\rm SM}(s) \, ds ~. \nonumber
\end{align}
When the above condition is met, it is also the case that $m_{\phi}R \gg 1$, implying that the last term above becomes subdominant. By contrast, the first two terms are well-approximated by integrating over a domain width $\delta r$ around $s = r$, where the exponential functions peak. Since the density varies slowly over the domain of integration, this yields
\begin{equation}
    \int_r^{r+\delta{r}} \frac{\exp(m_{\phi}(r-s))}{2m_{\phi}r} \, s \, n_{\rm SM}(s) \, ds \simeq \frac{1}{2m_{\phi}} n_{\rm SM}(r) \delta{r} + \mathcal{O}(\delta{r}^2)\,,
\end{equation}
\begin{equation}
    \int_{r-\delta{r}}^{r} \frac{\exp(-m_{\phi}(r-s))}{2m_{\phi}r} \, s \, n_{\rm SM}(s) \, ds \simeq \frac{1}{2m_{\phi}}  n_{\rm SM}(r) \delta{r} + \mathcal{O}(\delta{r}^2)\,.
\end{equation}

Combining the above expressions and approximating the domain width by $\delta{r} \simeq m_{\phi}^{-1}$, we arrive at 
\begin{equation}
    \phi_{\rm int}(r) \simeq - \frac{g_{\rm SM} n_{\rm SM}(r)}{m_{\phi}^2}~.
    \label{eq:phiint}
\end{equation}

The evaporation barrier, \textit{i.e.} the potential energy that a dark matter particle has from this additional field in the celestial object, is then obtained from the product $\phi_{\rm barrier}(r) = g_{\chi} \phi(r)$. This is the same scaling we estimated based on heuristic arguments in Eq.~\eqref{eq:approxpotential} in the main text (see also Ref.~\cite{Smirnov:2019cae} for a similar expression in the context of neutrino oscillations). 
Although our analytic solution given by Eqs.~\eqref{eq:app-a-mainsol-int} and \eqref{eq:app-a-mainsol-ext} is valid for any finite mediator mass, the simplified form in Eq.~(\ref{eq:phiint}) will hold for mediator masses $m_{\phi} \gtrsim 10^{-13} \ \rm eV$ for the celestial objects considered here. In the context of stellar cooling, Ref.~\cite{DeRocco:2020xdt} uses a similar scaling to Eq.~(\ref{eq:phiint}) in their estimates of stellar cooling limits with additional mediators that shift the mass of the radiated dark particles; our Eqs.~\eqref{eq:app-a-mainsol-int} and \eqref{eq:app-a-mainsol-ext} provide the generalized forms. See Ref.~\cite{Berlin:2023zpn} for a different trapping effect on positively charged millicharge particles in Earth, due to the electromagnetic charge in the Earth's atmosphere.

We also comment on the potential impact of field self-interactions on the barrier effect (see Ref.~\cite{Denton:2023iaa} for a detailed numerical analysis of the Klein-Gordon equation with quartic interactions). In the scalar case, a quartic potential $\sim \lambda \phi^4$ is generically expected since it can be generated through box diagrams with fermions running in the loop. We have verified by evaluating $\phi_{\rm barrier} = g_\chi \phi_\lambda(r)$, where $\phi_\lambda(r)$ is the full solution including a quartic potential and associated scalar mass corrections, that the condition $|\phi_{\rm barrier}(r = 0)| \gtrsim |\phi_{\rm grav}(r = 0)|$, where $\phi_{\rm grav}$ is evaluated at the evaporation mass in the absence of a barrier (see main text), is still met for couplings up to order unity in certain allowed regions of parameter space. This is because the inclusion of the dark matter density contribution in the full solution $\phi_\lambda(r)$ strengthens the barrier effect, compensating the screening produced by the quartic interaction. Therefore realizing our evaporation barrier does not require fine-tuning. Furthermore, we emphasize that the parameter space for which order one quartic couplings are allowed becomes larger for celestial objects residing in denser DM environments. This is because the full solution $\phi_\lambda(r)$ also receives a contribution from the interior DM density, which is considerably enhanced in higher DM densities. On the other hand, vector mediators can receive corrections from Euler-Heisenberg terms, which scale as $\sim g_\chi^4 F^{\mu \nu} F_{\mu \nu} / m_\chi^4$. In a similar manner, we have evaluated the above barrier condition in light of these corrections, and find that they do not generically impact our estimates. Cubic self-interactions can be analyzed in a very similar way. We note, however, that they are even more model dependent. In short, in both scalar and vector cases our scenario is viable without relying on fine-tuning assumptions.

\section{Dark Matter Effective Potential from Light Degrees of Freedom}
We now show how light fields sourced by celestial objects translate into an effective potential for the DM which, within a certain range of parameters, can prevent the evaporation of DM below the canonical mass values commonly assumed. We consider two benchmark cases, in which the DM has either a scalar or vector portal to the SM. Specific models that could generate such portals are discussed in App.~\ref{app:models} below. The central point we discuss here is that, in both cases, the fields sourced by celestial objects modify the dispersion relation of DM particles. In the low energy limit, this translates into a potential that increases the escape energy required for DM to evaporate.

If the DM is coupled to the SM via a real scalar $\phi$ with coupling strength $g_\chi$, the in-medium field inside the celestial object shifts the mass to an effective value
\begin{equation}
    m_{\chi}^{*}=m_{\chi}+ g_\chi \, \phi(\mathbf{r})~.
\end{equation}
Conservation of energy then implies that a DM particle with initial momentum $\mathbf{p}$ must be boosted to a momentum $\mathbf{p'}$ while transiting through due to the field sourced by the baryons, according to the relation
\begin{equation}
    \mathbf{p}^2+m_{\chi}^2 = \mathbf{p'}^2 + m_{\chi}^{* \, 2}~.
\end{equation}
In the limit that the effective mass shift is small compared to the bare DM mass $|g_\chi \phi(\mathbf{r})| \ll m_{\chi}$, the above expression can be expanded so that the kinetic energy change is expressed in terms of a non-relativistic potential
\begin{equation}
    \frac{p^2 - p'^2}{2m_{\chi}} \simeq g_\chi \, \phi(\mathbf{r})\equiv \phi_{\rm barrier}(\mathbf{r})~.
\end{equation}
See $e.g.$ Refs.~\cite{Acevedo:2020avd,Acevedo:2021kly,Acevedo:2022gug} for a similar discussion in the context of composite DM searches.

Similarly, the time-like component of a vector field $V_0(\vec{r})$ in the static limit also acquires an expectation value. However, the field shifts the total energy of the DM particle instead of its mass
\begin{equation}
    \sqrt{\mathbf{p}^2+m_{\chi}^2} = \sqrt{\mathbf{p'}^2 + m_{\chi}^2}- g_\chi \, V_0(\mathbf{r})~.
\end{equation}
As before, as long as $|g_\chi V_0(\mathbf{r})| \ll m_{\chi}$ and the DM particles are non-relativistic, both sides can be expanded to obtain
\begin{equation}
    \frac{p^2 - p'^2}{2m_{\chi}} \simeq g_\chi V_{0}(\mathbf{r})\equiv \phi_{\rm barrier}(\mathbf{r})\,.
\end{equation}
In this non-relativistic limit, vector fields yield an effective potential for DM particles of the same functional form as the scalar case for the celestial objects we consider. Note that for stronger interactions, the above treatment becomes non-trivial, and each case has to be treated separately in a quantum field theory framework. However, the field sourced by celestial objects is proportional to both the coupling and number density of SM carriers. Since large couplings are quite constrained by experiments, such a relativistic regime would be inaccessible to most celestial objects unless they have extreme densities to compensate for the small coupling, such as white dwarfs and neutron stars.

\section{Particle Model Realizations for the Evaporation Barrier}
\label{app:models}

As detailed in the previous section, both scalar and vector portals can generate an additional potential within celestial objects that prevents light DM from evaporating. Here, we discuss some well-motivated models for which this effect can be prominent, given existing constraints on new light mediators within the SM sector. In all cases, we assume that the DM is a Dirac fermion $\chi$, although Majorana fermions or bosonic DM scenarios could be explored as well.
\begin{itemize}
    \item \textit{Light Scalar Mediators:} A real light scalar generically mediates an attractive force between SM nucleons and the DM, with interactions of the form
      \begin{equation}
       \mathcal{L}\supset \left(g_\chi\bar{\chi}\chi+g_n\bar{n}n\right)\phi - \frac{1}{2} m_{\phi}^2\phi^2.
       \label{eq:app-b-samplemodel1}
      \end{equation}
  The coupling to nucleons can be realized through a coupling of the scalar to the top quark or a colored vector-like generation~\cite{Knapen:2017xzo}. Alternatively, one could consider a similar construction for a leptophilic scalar mediator that couples the DM to the electron field~\cite{Parikh:2023qtk}, though the leptophilic scenario is more constrained by BBN and stellar cooling.  Overall, models with such light particles have been considered extensively in the literature, and are a well-motivated field of SM extensions. For example, they have been considered in the context of nuclear decay observations~\cite{Feng:2016ysn,Fornal:2017msy}, or the neutron lifetime anomaly~\cite{Fornal:2019olr}. 

  If self-interactions of the mediator were included for such scalar mediator models, quartic self-couplings could induce in-medium mass corrections of the order $\delta m_\phi^2 \sim \lambda g_{\rm SM}^2 n_{\rm SM}^2 / m_{\phi}^4$. If the self-interaction is chosen to be repulsive, this could potentially weaken the evaporation barrier for ultralight mediators unless the self-coupling $\lambda$ is small at tree-level. In addition, thermal corrections to the scalar mass are possible but are of order $\delta m_{\phi}^2 \sim g_{\rm SM} n_{\rm SM}/T$, and are therefore negligible for the parameter space considered.

\item \textit{Axion-like particles:} A particularly compelling candidate for a light particle is the axion, which is introduced as a solution to the strong CP-problem~\cite{Peccei:1977hh,Adams:2022pbo}. The relevant coupling to the QCD sector arises from the axion-gluon interaction
\begin{align}
\mathcal{L} \supset \frac{a}{f_a} \, \frac{\alpha_s}{8 \pi} G_{\mu \nu}^a \tilde{G}^a_{\mu \nu}\,,
\end{align}
where $a$ is the axion field, $f_a$ the axion decay constant $G^a_{\mu\nu}$ the gluon field tensors, and $\alpha_s$ the strong coupling constant. It has been discussed in Refs.~\cite{Bigazzi:2019hav,Okawa:2021fto,Fan:2023hci} that in the presence of CP-violation, which can either be the effect of an imperfect Peccei-Quinn symmetry or transferred from the electroweak sector of the SM, a CP-odd $g_{a NN}$ coupling can arise leading to a non-derivative axion-nucleon interaction. The expected coupling strength needed to have a sizable effect on evaporation is compatible with the commonly assumed mass and coupling range of common axion scenarios~\cite{Shifman:1979if,Dine:1981rt}, as discussed in Ref.~\cite{Bigazzi:2019hav}. Similarly, the axion may couple to DM %$g_\chi a \, \bar{\chi} \chi$, 
with dark sector coupling $g_\chi$, which is only weakly constrained from the phenomenological perspective. The relevant interaction Lagrangian is thus analogous to the above scenario
\begin{align}
\mathcal{L} \supset g_{a NN} a \bar{N} N  + g_\chi a \, \bar{\chi} \chi - \frac{1}{2} m_a^2 a^2\,. 
\end{align}
 The axion is expected to induce a long-range force between the DM and SM particles when its mass is light, as is commonly assumed in the literature.  The parameter space for this long-range interaction is analogous to the one shown in Fig.~\ref{fig:evapthreshold}. 
 
 Note that in the case of the axion the quartic self-coupling is naturally small, since it scales as $\lambda_a \sim m_a^2/f_a^2$~\cite{GrillidiCortona:2015jxo}, where $f_a$ is typically required to be above $\sim 10^{12}$ GeV. Thus in-medium mass corrections are negligible in this scenario.
 
    \item \textit{New $U(1)$ gauge bosons:} Numerous BSM theories have considered baryon and lepton numbers or a combination of these as new gauge symmetries that are spontaneously broken \cite{FileviezPerez:2010gw,Bell:2014tta,Tulin:2014tya,Dobrescu:2014fca}. 
    To illustrate the applicability of this scenario, we consider a new $U(1)_B$ gauge interaction which can be parametrized at the nucleon level as
     \begin{equation}
         \mathcal{L} \supset \left(g_\chi\bar{\chi}\gamma^{\mu}\chi+g_n\bar{n}\gamma^{\mu}n\right) V_{\mu}-\frac{1}{2}m_V^2 V_{\mu}V^{\mu},
     \end{equation}
     where the time-like component acquires a non-zero expectation value inside the celestial object in the static limit. For the above model, additional field content is generally needed to ensure it is anomaly-free~\cite{FileviezPerez:2010gw, Duerr:2013dza, Duerr:2017whl}, and gauge invariant ~\cite{Kors:2004dx,Feldman:2007wj,Duerr:2017whl} which we avoid detailing here. Theories with additional gauged quantum numbers have been considered extensively in a number of outstanding problems such as the $(g-2)_{\mu}$ anomaly \cite{Lee:2014tba,Altmannshofer:2016brv}, flavour violating decays \cite{Dutta:1994dx,Heeck:2011wj,Altmannshofer:2016oaq}, DM \cite{Duerr:2013lka,Schwaller:2013hqa,Batell:2014yra,Bell:2016fqf, Bell:2016uhg, Bell:2017irk} and neutrino mass generation \cite{Lindner:2011it,Kanemura:2014rpa}.
\end{itemize}

\begin{figure}[t!]
    \centering
     \includegraphics[width=0.382\linewidth]{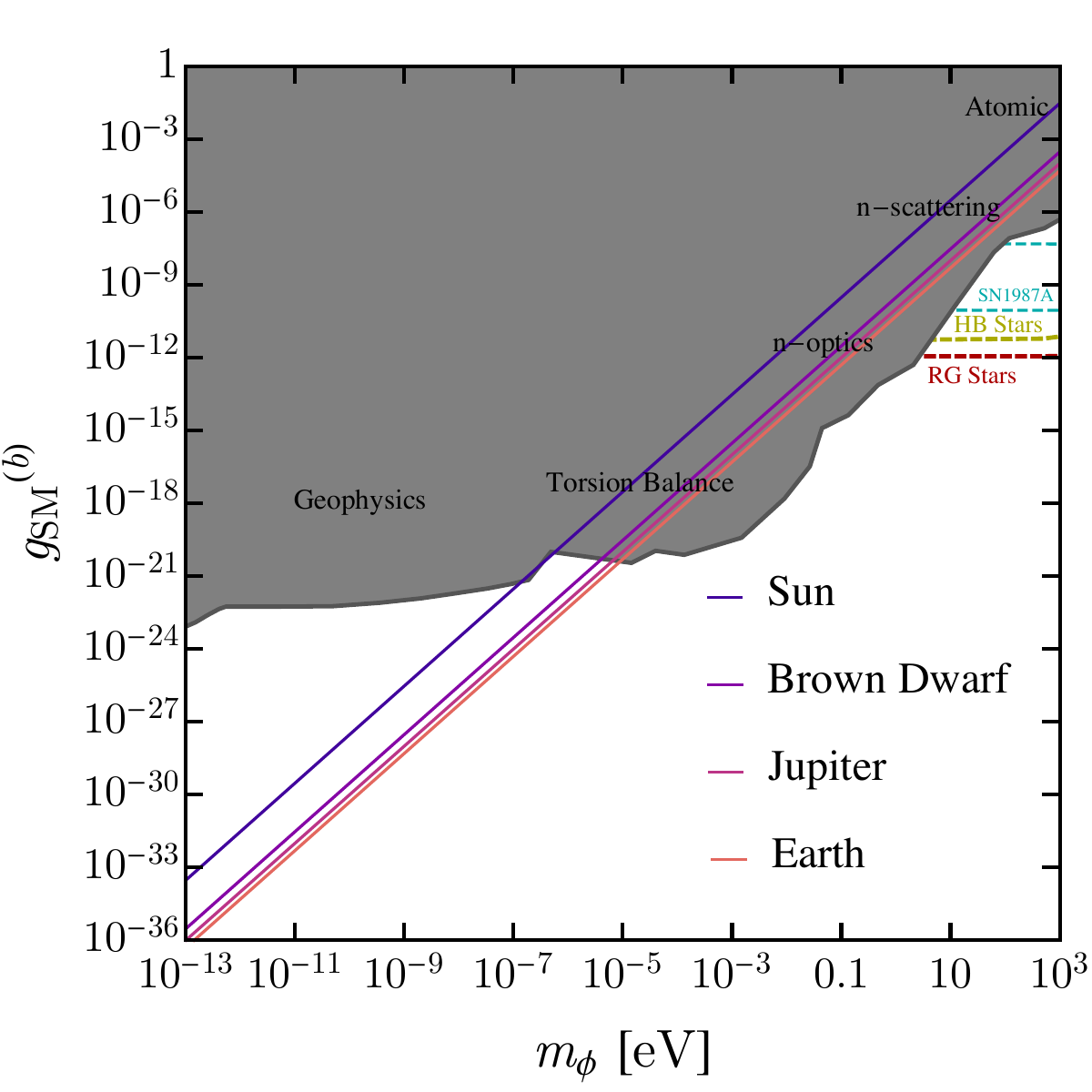}
    \includegraphics[width=0.382\linewidth]{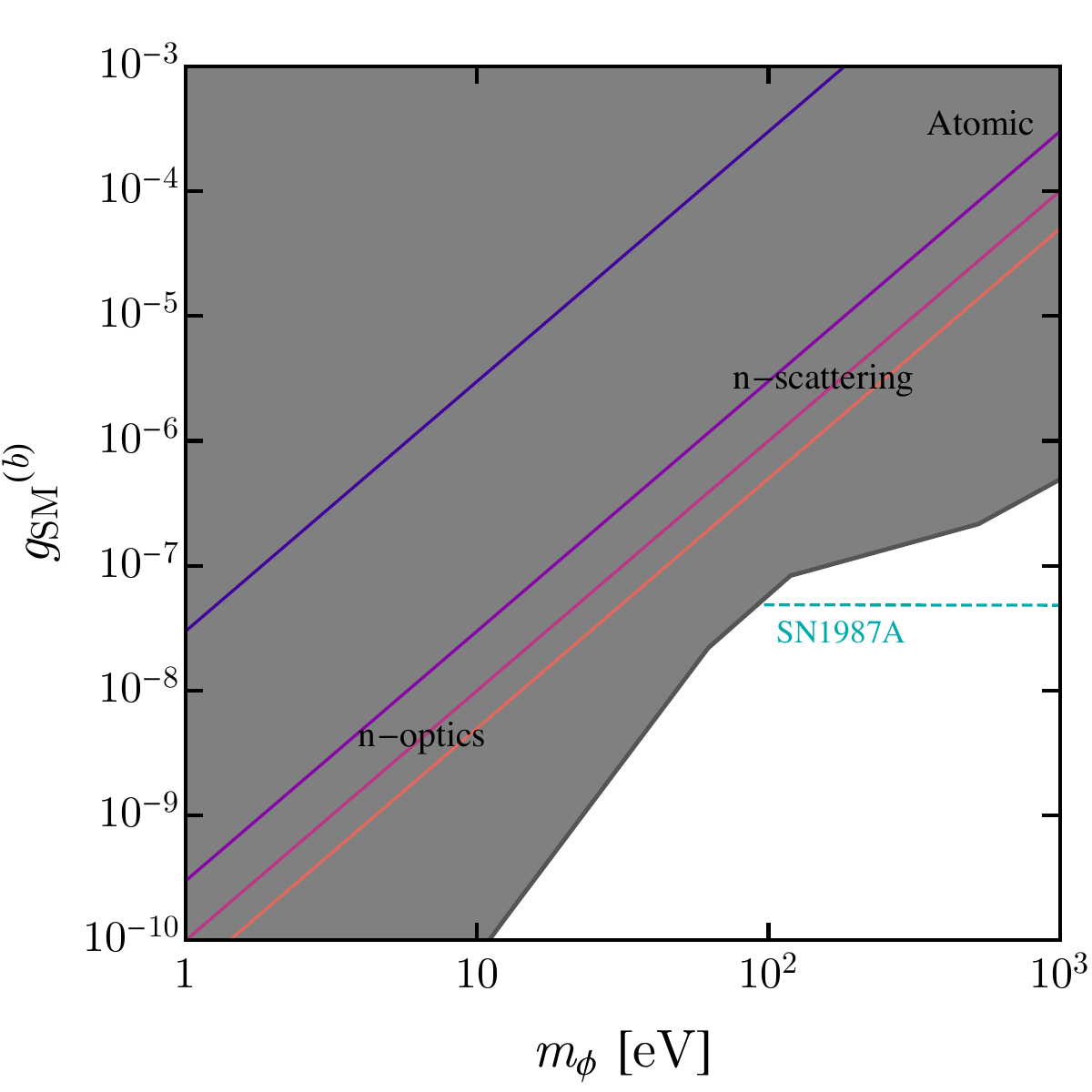}\\
     \includegraphics[width=0.382\linewidth]{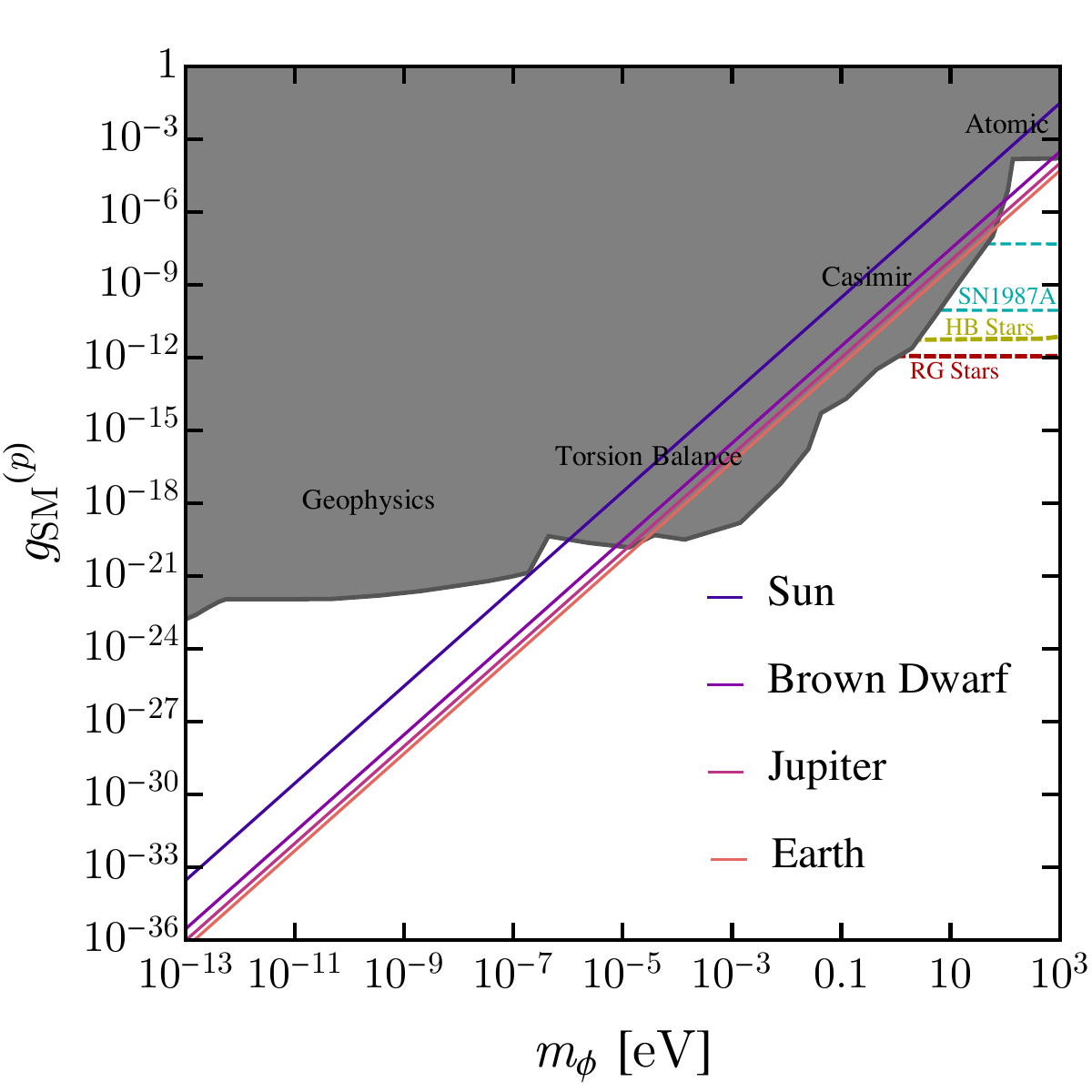}
          \includegraphics[width=0.382\linewidth]{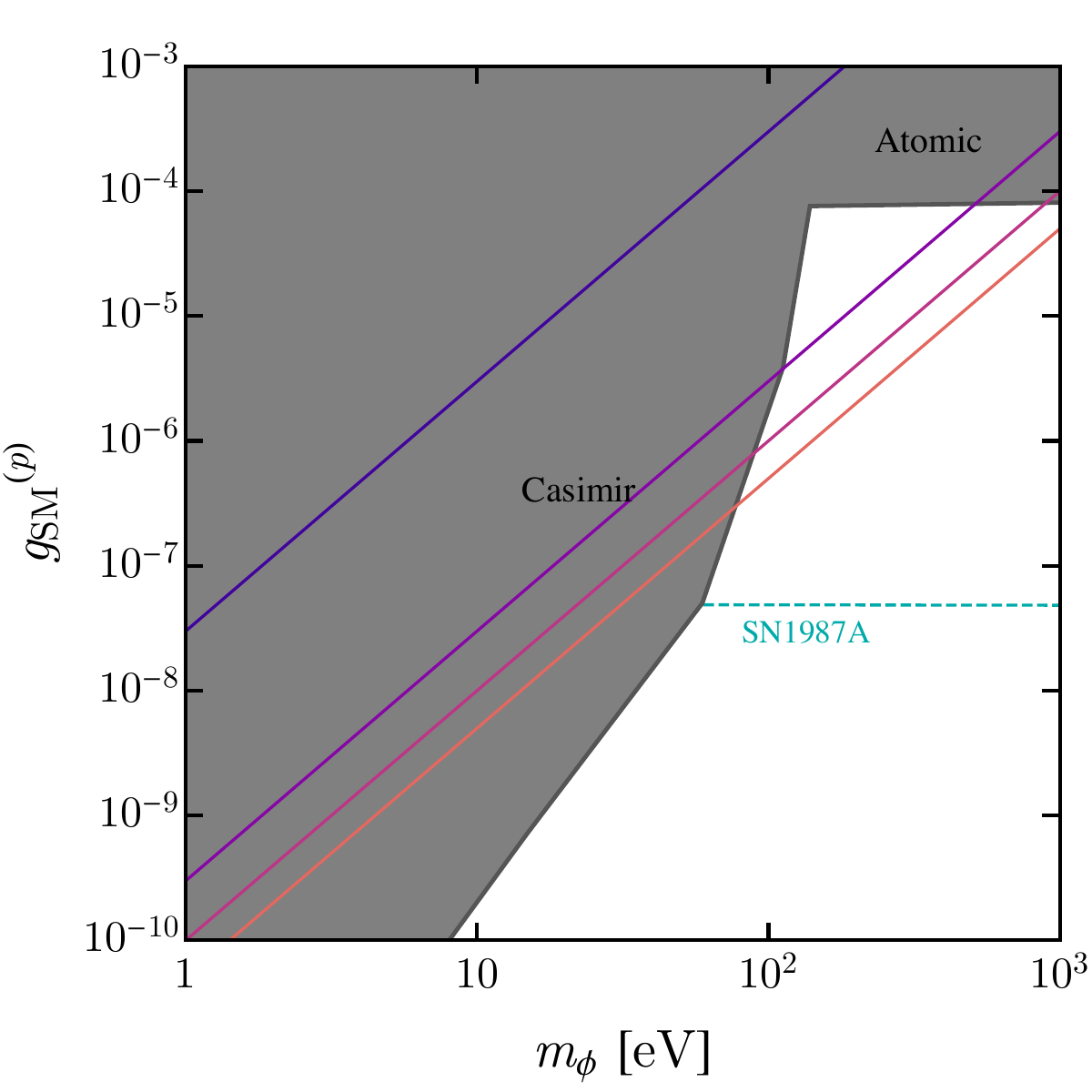}\\
     \includegraphics[width=0.382\linewidth]{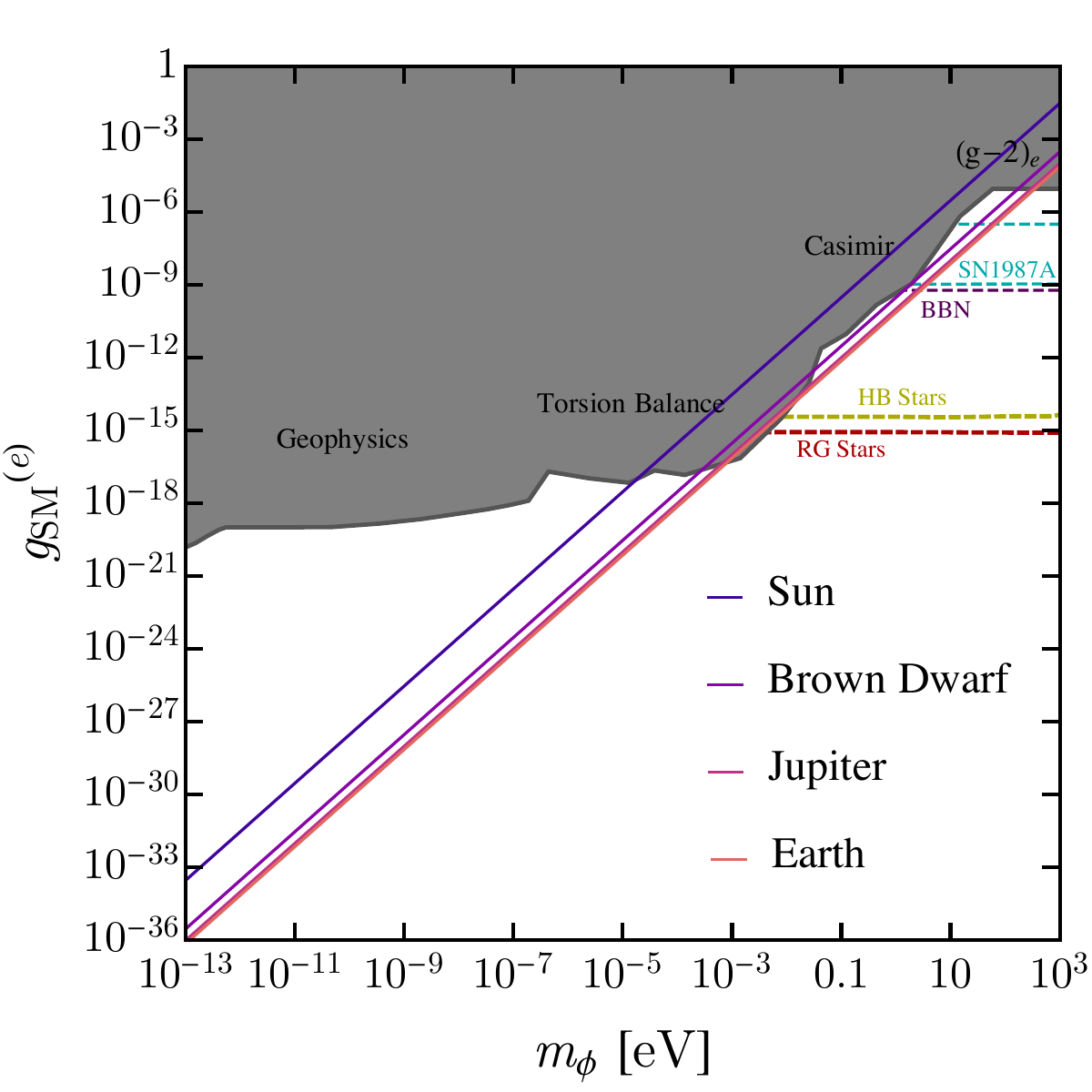}
          \includegraphics[width=0.382\linewidth]{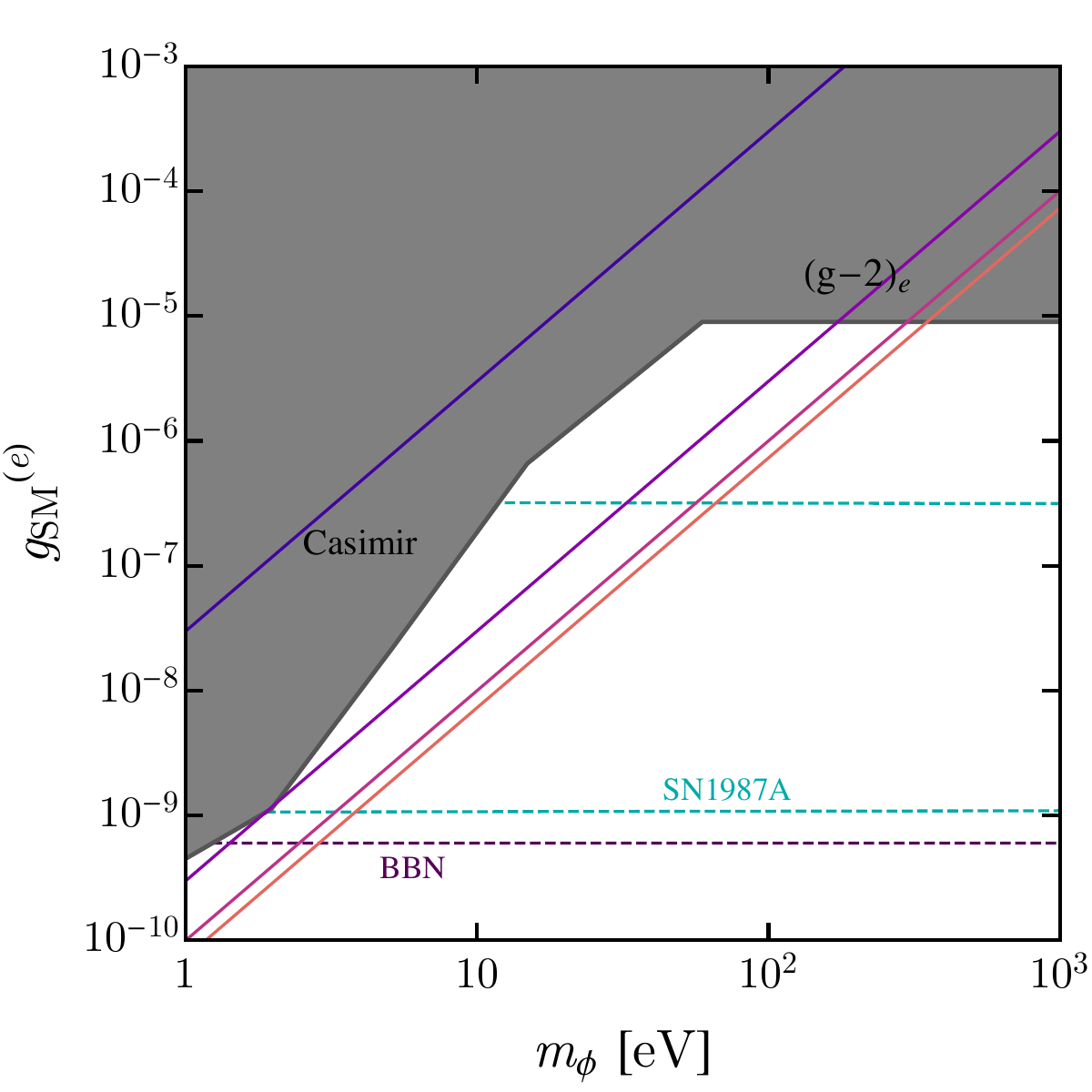}

    \caption{Constraints on long-range mediators with equal coupling to protons and neutrons (\textbf{top row}), proton-only coupling (\textbf{middle row}), and electron-only coupling (\textbf{bottom row}); see text for constraint details. Overlaid we show minimum coupling values $g_{\rm SM}$ (assuming $g_\chi=1$) for which the resulting evaporation barrier dominates in each celestial object, and therefore reduces the DM evaporation mass. The right column is simply a zoom in on the high mediator mass region of the left column.}
    \label{fig:alpha-supp}
\end{figure}

Figure~\ref{fig:alpha-supp} illustrates the minimum coupling $g_{\rm SM}$ for which evaporation barrier effects dominate in each celestial object, compared to existing experimental constraints. To facilitate straightforward comparison, we have set the DM coupling $g_\chi = 1$. We show three coupling cases on each row: equal proton and neutron couplings, proton couplings, or electron couplings. These couplings are partly constrained by experiments including torsion balance \cite{Hoskins:1985tn,Hoyle:2004cw,Adelberger:2006dh,Schlamminger:2007ht,Sushkov:2011md,Yang:2012zzb}, casimir-type searches \cite{Lamoreaux:1996wh,Obrecht:2006fqr,Masuda:2009vu,Bezerra:2010pq,Bezerra:2011xc}, atomic and nuclear measurements \cite{Xu:2012wc,Murata:2014nra} and geophysical data \cite{Adelberger:2003zx}. For isospin-invariant couplings, there are additional constraints from neutron optics and neutron-xenon scattering experiments \cite{Leeb:1992qf,Nesvizhevsky:2004qb,Greene:2006qj,Baessler:2006vm,Nesvizhevsky:2007by}. For proton-only couplings, we have rescaled the limits by a factor $Z/A \simeq 1/2$ and neglected the mass difference between neutrons and protons. Similarly, for electron-only couplings, we have included $(g-2)_e$ constraints \cite{Liu:2016qwd} and rescaled other applicable limits form 5$^{\rm th}$ force searches by a factor $m_e / m_N \simeq 1/1836$. 
We also show astrophysical constraints for scalar mediators based on stellar cooling of RG and HB stars \cite{Hardy:2016kme} (see also Ref.~\cite{Bottaro:2023gep} for similar constraints from white dwarfs), and SN1987A \cite{Rrapaj:2015wgs,Chang:2018rso}, as well as cosmological constraints based on BBN deuterium abundances for leptophilic mediators \cite{Cyburt:2015mya,Knapen:2017xzo}. For vector mediators, these limits are in general weaker in the light mass end than the scalar bounds shown in Fig.~\ref{fig:alpha-supp}. While we show astrophysical bounds as dashed lines for completeness, it is important to keep in mind that they are highly model dependent, and can disappear under a similar mechanism to the evaporation barrier, as discussed in Ref.~\cite{DeRocco:2020xdt}. In addition to the previous limits, black hole superradiance places coupling-independent constraints on light bosonic particles in mass windows below $\lesssim 10^{-11} \ \rm eV$ \cite{Brito:2015oca,Cardoso:2018tly,Baryakhtar:2017ngi,Baryakhtar:2020gao}. However, these limits crucially depend on black hole spin measurements~\cite{Arvanitaki:2010sy}, for which there is not a consensus~\cite{Mathur:2020aqv}. Simple model building \cite{Mathur:2020aqv} and back-reaction effects \cite{Blas:2020kaa}, for example, can entirely lift these constraints in some cases.

Overall, we emphasise that while data from BBN and CMB have potential to constrain light dark sectors, a crucial required assumption is that those sectors are thermalized during the relevant epochs. Several scenarios avoid those bounds all together. One possibility is that the dark sector never enters thermal and kinetic equilibrium with the SM, and the DM abundance is produced non-thermally. Another possibility is that a late-time phase transition makes light degrees of freedom light only at late times~\cite{Elor:2021swj}. Or as a third option,  thermalization can only take place after a late-time phase transition~\cite{Parikh:2023qtk}. In all these scenarios dark sectors with light particles are consistent with cosmological observations.

Finally, we also comment on constraints arising from DM self-interactions, which can generically arise in some of these models. Observations of the Bullet Cluster and halo shapes have led to claimed limits on DM self-couplings through light mediators of order $g_{\chi} \lesssim 10^{-4} \ (m_\chi/{\rm MeV})^{3/4}$ for velocities that would be relevant for captured DM in celestial objects \cite{Knapen:2017xzo}. However, we emphasize that these bounds are not robust, due to their dependence on how observations of the astrophysical probe are performed~\cite{Robertson:2016xjh,Robertson:2022pjy,Popesso_2006, Wittman_2018,Adhikari:2022sbh}, and a lack of systematic error bars. They also depend on the DM model assumptions~\cite{Knapen:2017xzo}. In any case, even taking these bounds at face value, ample viable parameter space exists where the evaporation barrier present in celestial objects could trap light DM.

\section{Behaviour of Evaporation Contours}
\label{app:escapes}

\begin{figure*}[!th]
    \centering          \includegraphics[width=0.45\linewidth]{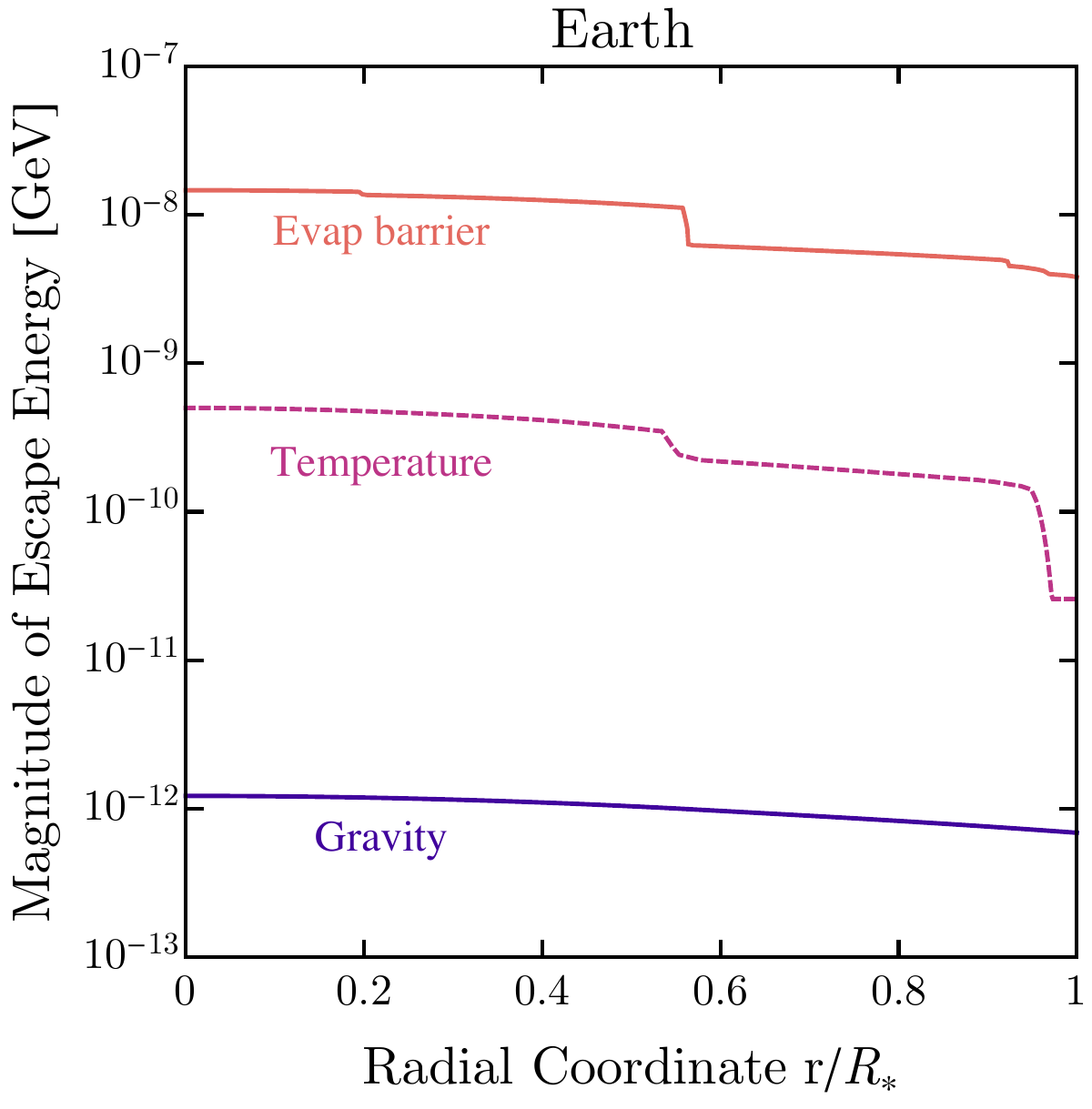}\hspace{5mm}
     \includegraphics[width=0.45\linewidth]{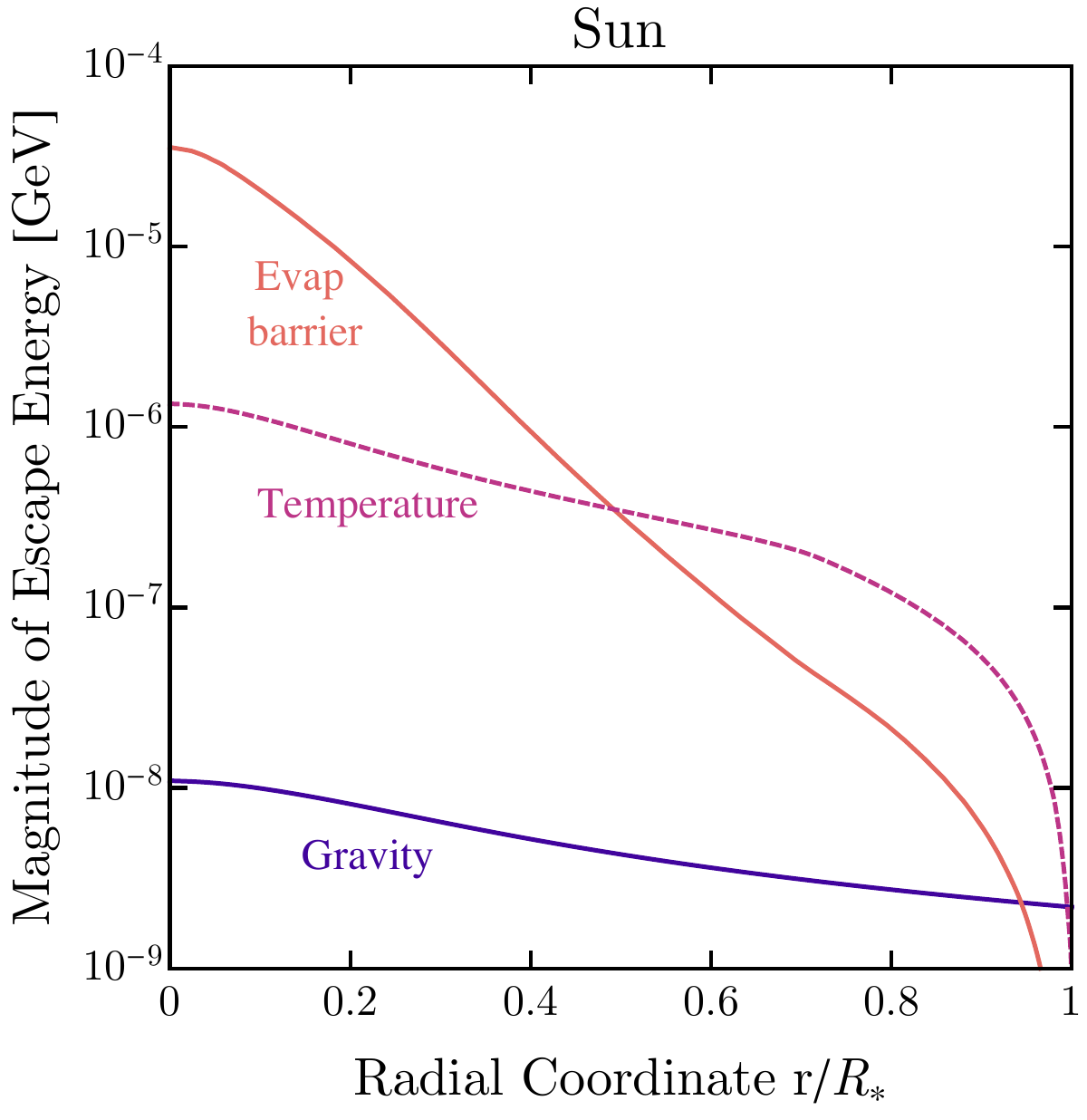}\vspace{5mm}
    \includegraphics[width=0.45\linewidth]{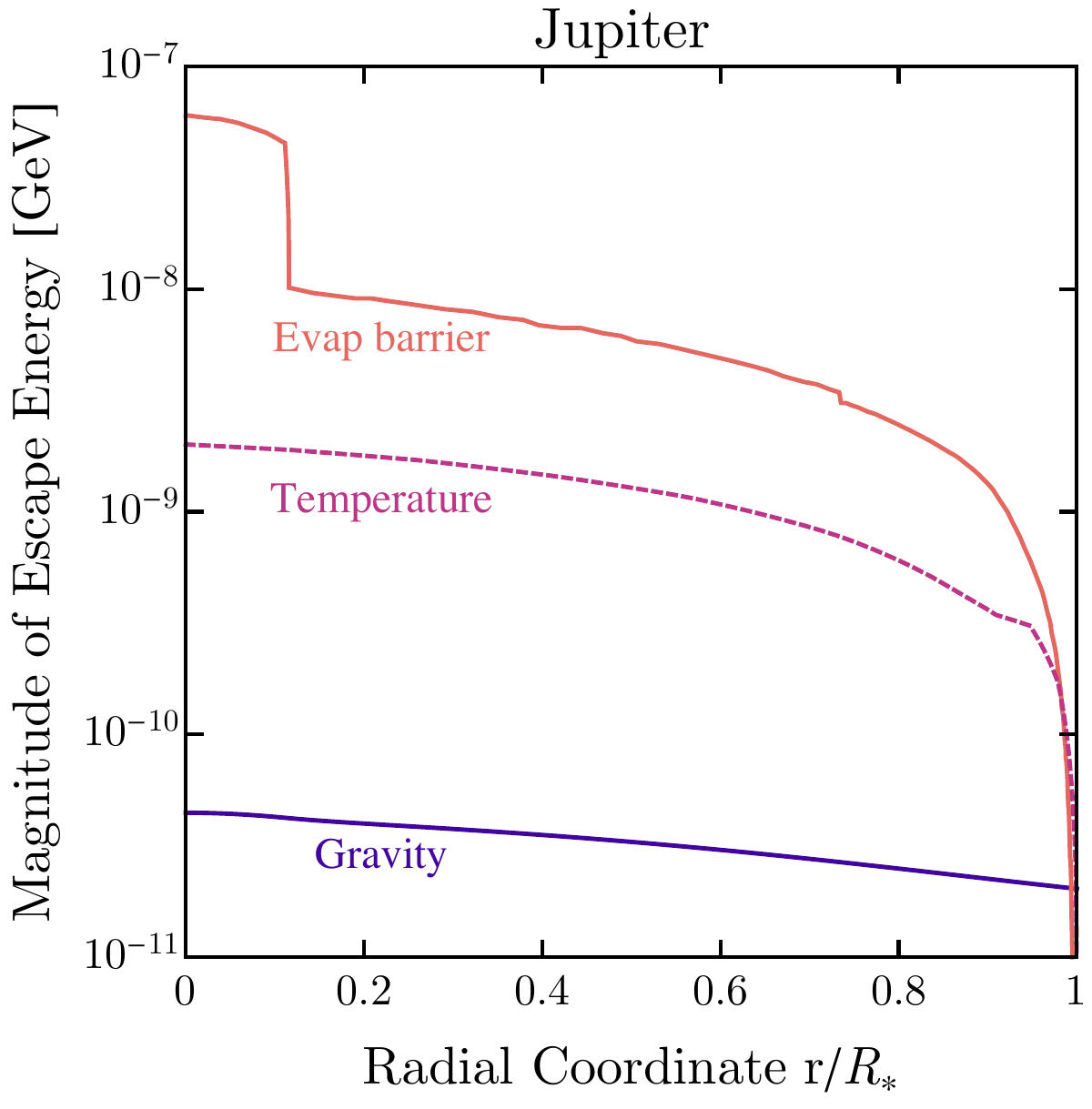}\hspace{5mm}
     \includegraphics[width=0.45\linewidth]{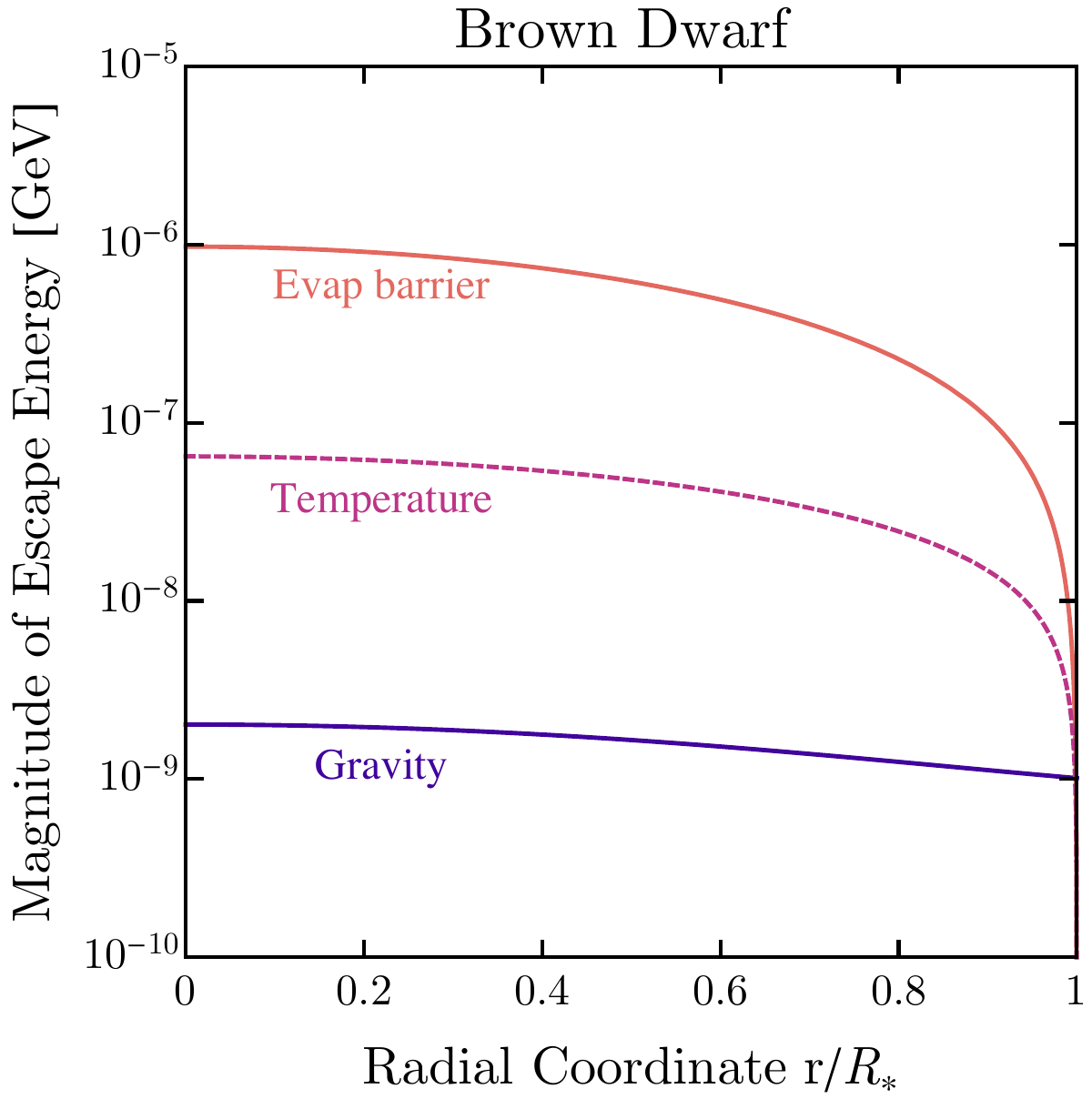}
    \caption{   
    Magnitude of the total energy needed to escape from radius $r$ for a 1 MeV DM particle due to gravity (solid) and the evaporation barrier (solid), for each celestial object as labeled. Here we have chosen the new long-range force couplings such that $\phi_{\rm barrier} \gg \phi_{\rm grav}$ at the core of the object. For comparison, we include the temperature (dashed) as a function of radius $r$ which is a measure of the local evaporation efficiency.
    }
    \label{fig:escapes}
\end{figure*}

\subsection{Dependence on Cross Section and DM Mass}

There are several factors that lead to evaporation or retention of DM in celestial objects. First, higher or lower ambient temperatures lead to larger or smaller thermal kicks imparted to the DM. Second, the size of the DM-SM scattering rate determines how frequently the DM receives thermal kicks from the ambient SM temperature, and it is also relevant for keeping the DM trapped by down-scattering the thermally kicked DM. Third, the amount of celestial-body overburden available to keep the DM trapped is relevant -- if there are too few SM particles standing between the DM and the surface for a given interaction rate, the DM will not be retained. Fourth, the lighter the DM is, the easier it is for it to escape the gravitational well. These features appear in our Fig.~\ref{fig:evapcontours} in the main text. The lowest evaporation masses occur in the two extremes: either the strong interaction limit, or weak interaction limit. In the strong limit, DM is thermally kicked often, but it is also quickly down-scattered again, due to the large optical depth between the DM and the surface. In the weak limit, DM receives very few thermal kicks, so the few down-scatters do not matter given DM is not excited much in the first place. The highest evaporation masses usually occur at the cross sections roughly where DM interactions switch from the weak to the strong interaction regime. This cross section is about $5 \times 10^{-36}$~cm$^2$, $10^{-36}$~cm$^2$, $10^{-35}$~cm$^2$, and $2 \times 10^{-36}$~cm$^2$ for the Earth, Sun, Jupiter, and our benchmark Brown Dwarf, respectively. In this case, the interaction is not so strong that the thermally kicked DM can be scattered back down and remain trapped, but it is not so weak that the thermal kicks happen so infrequently that evaporation occurs less often. 

The optimal celestial-body radius for evaporation within the celestial object varies strongly as a function of the DM-SM scattering rate. In fact, for high cross sections, the DM evaporation occurs only in a very thin shell towards the celestial-body surface~\cite{Gould:1989hm}. For smaller cross sections, evaporation optimally occurs closer to the core~\cite{Gould:1989hm}.

Our additional evaporation barrier effect serves to re-enforce gravity, offering an additional potential that also pulls the DM particles towards the core. The shape of the curves we see in Fig.~\ref{fig:evapcontours} of the main text is dominantly for the reasons we just outlined above, though there are some interesting effects we observe which do not occur in the gravity only case, which we now discuss below.

\subsection{Comparison of Escape Energy to Temperature in Celestial Objects}

Figure~\ref{fig:escapes} compares the magnitude of the escape energy due to the gravitational field or new long-range force (which traces the SM density profile of the object), to the temperature, for each celestial object. For concreteness, we have set the DM mass to 1 MeV, which lies below the evaporation mass for these objects when considering only the gravitational force. For the Earth, Jupiter, or the Brown Dwarf, we see that when the long-range force dominates at the core of the object, it also dominates at the surface. Given that at high cross-sections the DM evaporation occurs in a very thin shell at the surface~\cite{Gould:1989hm}, this means that even at high cross sections the new long-range force will dominate in these objects and prevent DM evaporation. However, this is not true for the Sun. In this case, due to the shape of the solar SM density profile traced by the new long-range force, compared to the solar temperature profile, we see that while the new long-range force can dominate at the core, it will not dominate at radii closer to the solar surface unless the coupling is significantly stronger compared to the other cases. This means that, for sub-GeV DM masses in the Sun, while the new long-range force blocks evaporation for lower cross sections (which have evaporation shells closer to the core), it will not generically block evaporation at higher cross sections (which have evaporation shells closer to the surface). This is exactly what we observe for the Sun in our main text Fig.~\ref{fig:evapcontours}, where at high cross sections the reduction in the DM evaporation mass is not as prominent compared to lower cross sections.

Comparing instead just the gravitational escape energy and the temperature in Fig.~\ref{fig:escapes}, we note that temperature is roughly dominating everywhere over gravity for all the objects. For this reason, without including our evaporation barrier, DM below about 250 MeV with spin-independent dominated capture and evaporation interactions was previously claimed to evaporate in all celestial objects~\cite{Garani:2021feo}. We have however shown that whether or not DM below about 250 MeV evaporates is highly-model dependent. In addition, note that the fitting procedure for brown dwarf core temperatures in Ref.~\cite{Garani:2021feo} extrapolates results of Ref.~\cite{Paxton:2010ji} that go to 3 Gyr, resulting in higher brown dwarf core temperatures compared to core temperatures at 10 Gyr~\cite{Leane:2020wob} where an analytic brown dwarf evolution model is used~\cite{2016AdAst2016E..13A}.
This overall leads to higher DM evaporation mass estimates for brown dwarfs in Ref.~\cite{Garani:2021feo} even in the gravity only case.

\bibliography{dwarfs}	
\end{document}